\documentclass[sigconf,nonacm]{acmart}
\usepackage{xcolor}
\usepackage[utf8]{inputenc}
\usepackage{amsmath}
\usepackage{amsfonts}
\usepackage{graphicx}
\usepackage{siunitx}
\usepackage{array}
\usepackage{listings}
\usepackage{soul}
\usepackage{mathtools}
\usepackage{tikz}
\usetikzlibrary{decorations.pathreplacing}
\usepackage{float}
\usepackage{tabularx}
\usepackage{algorithm}
\usepackage{algpseudocode}
\newcolumntype{Y}{>{\centering\arraybackslash}X}
\usepackage{balance}
\usepackage{hyperref}
\usepackage{url}
\usepackage{xspace}
\usepackage{tcolorbox}
\usepackage{enumitem}
\usepackage{subcaption}
\usepackage{multirow}
\usetikzlibrary{shapes}
\usetikzlibrary{patterns, fit}
\usepackage{adjustbox}
\usepackage{pifont}
\usepackage{changepage}
\usepackage[normalem]{ulem}

\acmConference[ISSTA 2024]{ACM SIGSOFT International Symposium on Software Testing and Analysis}{16-20 September, 2024}{Vienna, Austria}

\newcommand{\dcircle}[1]{\ding{\numexpr171 + #1}}

\usepackage[frozencache]{minted}
\setminted[java]{
xleftmargin=15pt,
framesep=2mm,
linenos=true,
breaklines=true,
tabsize=2,
encoding=utf8,
frame=none,
fontsize=\footnotesize,
escapeinside=|| 
}
\usepackage{cleveref}

\newcommand{\highlight}[1]{%
\begin{tcolorbox}[leftrule=1mm,rightrule=1mm,toprule=0mm,bottomrule=0mm,left=0pt,right=0pt,top=0pt,bottom=0pt, colback=gray!30, colframe=gray!90]
#1
\end{tcolorbox}%
}

\makeatletter
\newcommand{\ostar}{\mathbin{\mathpalette\make@circled\star}}
\newcommand{\make@circled}[2]{%
  \ooalign{$\m@th#1\smallbigcirc{#1}$\cr\hidewidth$\m@th#1#2$\hidewidth\cr}%
}
\newcommand{\smallbigcirc}[1]{%
  \vcenter{\hbox{\scalebox{0.97778}{$\m@th#1\bigcirc$}}}%
}
\newbox\dottedarrow@box
\setbox\dottedarrow@box\hbox
  {%
    \begin{tikzpicture}
      \draw[dotted,->] (0,0) -- (1.5em,0);
    \end{tikzpicture}%
  }
\newcommand*\dottedarrow
  {\relax\ifmmode\expandafter\dottedarrow@m\else\expandafter\dottedarrow@t\fi}
\newcommand*\dottedarrow@t[1][1.5em]
  {\resizebox{#1}{!}{\raisebox{.5ex}{\usebox\dottedarrow@box}}}
\newcommand*\dottedarrow@m[1][]
  {%
    \if\relax\detokenize{#1}\relax
      \mathchoice
        {\dottedarrow@t}
        {\dottedarrow@t}
        {\dottedarrow@t[1.1em]}
        {\dottedarrow@t[0.9em]}%
    \else
      \dottedarrow@t[#1]%
    \fi
  }
  
\algnewcommand\algorithmicswitch{\textbf{switch}}
\algnewcommand\algorithmiccase{\textbf{case}}
\algdef{SE}[SWITCH]{Switch}{EndSwitch}[1]{\algorithmicswitch\ #1\ \algorithmicdo}{\algorithmicend\ \algorithmicswitch}%
\algdef{SE}[CASE]{Case}{EndCase}[1]{\algorithmiccase\ #1}{\algorithmicend\ \algorithmiccase}%
\algtext*{EndSwitch}%
\algtext*{EndCase}%
  
\makeatother

\newcolumntype{g}{>{\columncolor{gray!30}}l}
\newcolumntype{h}{>{\columncolor{gray!30}}c}

\def\BibTeX{{\rm B\kern-.05em{\sc i\kern-.025em b}\kern-.08em
    T\kern-.1667em\lower.7ex\hbox{E}\kern-.125emX}}

\newcommand{\mathid}[1]{\ensuremath{\mathit{#1}}}
\def\|#1|{\mathid{#1}}
\protected\def\codeid#1{\ifmmode{\mbox{\smaller{\ttfamily{#1}}}}\else{\smaller{\ttfamily #1}}\fi}
\def\<#1>{\codeid{#1}}

\newcommand{\numtools}{13\xspace}

\newcommand{\doInBackground}{\<do>\-\<In>\-\<Back>\-\<ground>\xspace}
\newcommand{\onPostExecute}{\<on>\-\<Post>\-\<Ex>\-\<e>\-\<cute>\xspace}

\begin{document}

\title{Call Graph Soundness in Android Static Analysis}

\author{Jordan Samhi}
\orcid{0000-0001-6052-6184}
\affiliation{%
  \institution{CISPA Helmholtz Center for Information Security}
  \city{Saarbr{\"u}cken}
  \country{Germany}
}
\email{jordan.samhi@cispa.de}

\author{René Just}
\orcid{0000-0002-5982-275X}
\affiliation{%
  \institution{University of Washington}
  \city{Seattle}
  \country{USA}
}
\email{rjust@cs.washington.edu}

\author{Tegawendé F. Bissyandé}
\orcid{0000-0001-7270-9869}
\affiliation{%
  \institution{University of Luxembourg}
  \city{Luxembourg}
  \country{Luxembourg}
}
\email{tegawende.bissyande@uni.lu}

\author{Michael D. Ernst}
\orcid{0000-0001-9379-277X}
\affiliation{%
  \institution{University of Washington}
  \city{Seattle}
  \country{USA}
}
\email{mernst@cs.washington.edu}

\author{Jacques Klein}
\orcid{0000-0003-4052-475X}
\affiliation{%
  \institution{University of Luxembourg}
  \city{Luxembourg}
  \country{Luxembourg}
}
\email{jacques.klein@uni.lu}

\begin{abstract}
Static analysis is sound in theory, but an implementation may un\-sound\-ly fail to analyze all of a
program's code.  Any such omission is a serious threat to the validity of
the tool's output.  Our work is the first to measure the prevalence of these omissions.
Previously, researchers and analysts did not know what is
missed by static analysis, what sort of code is missed, or the reasons behind these omissions.
To address this gap, we ran \numtools static analysis tools and a dynamic
analysis on 1000 Android apps.
Any method in the dynamic analysis but not in a static analysis is an unsoundness.

Our findings include the following.
\dcircle{1} Apps built around external frameworks
challenge static analyzers. On average, the \numtools static analysis tools failed to capture 61\% of the dynamically-executed methods.
\dcircle{2} A high level of precision in call graph construction is a
synonym for a high level of unsoundness;
\dcircle{3} No existing approach significantly improves
static analysis soundness.  This includes those specifically
tailored for a given mechanism, such as DroidRA to address
reflection. It also includes systematic approaches, such as Edge\-Miner, capturing all callbacks in the Android framework systematically.
\dcircle{4}Modeling entry point methods challenges call graph construction which jeopardizes soundness.

\label{sec:dummy-for-etags}
\end{abstract}

\maketitle

\section{Introduction}
\label{sec:introduction}

A static analysis can be conservative and sound
by computing with a model that over-approximates the code's behavior.
In contrast, dynamic analysis has good precision, as it computes
concrete values and captures actual run-time behavior, but it
is unsound because it
knows nothing about other executions or about un-covered code.
Hence, analysis clients can choose soundness (static analysis) or precision (dynamic analysis)~\cite{Ernst2003:WODA}.
However, this is to some extent a false choice:  soundness is often not
available in practice due to unsound static models. This paper
  investigates this phenomenon in the context of analyzing Android apps.

One reason for unsoundness when analyzing Android apps is that
apps are event- and callback-driven, resulting in a complex
execution flow that static analysis struggles to capture~\cite{10.1109/ICSE.2017.36,10.1109/ASE.2019.00021}.
For instance, static analysis fails to fully and automatically account for the \emph{implicit} invocation of methods by frameworks, such as the Android framework, Flutter~\cite{flutter}, Xamarin~\cite{xamarin}, etc.
The calls to some application methods are in a
framework, which (for
scaling reasons) is often not statically analyzed along with
the app.
Missing these implicit calls leads to an incomplete understanding of the
app's behavior (e.g., missing nodes and edges in the call graph)~\cite{7372054,10.1145/2931037.2931044}.
This blind spot can be exploited by attackers to circumvent state-of-the-art static analysis tools, such as FlowDroid~\cite{10.1145/2666356.2594299}, and hide malicious code~\cite{scylla}.
Hence, developing the most effective static data leak detector or static
malware detector is little help if the analysis is run over
an unsound static model of the app. 

Numerous approaches have attempted to refine call graphs by accounting for
specific
mechanisms~\cite{10.1145/2666356.2594299,10.1145/3510003.3512766,BarrosJMVDdAE2015,10.1145/2931037.2931044,10.5555/2818754.2818791,10.1109/ICSE43902.2021.00126,gordon2015information,reunify},
such as reflection.
In addition, there have been attempts to systematically analyze the Android framework to collect callbacks~\cite{columbus,cao2015edgeminer}.
None of these approaches is comprehensive.
Although some studies show that several mechanisms, such as implicit calls, cause unsound static analysis~\cite{cao2015edgeminer,7372054}, the extent of under-approximation of static models is not known.
As a result, there is a need to systematically explore the amount of methods missed during static analysis and study their underlying causes.

Our work consists of three parts.  First, we identified
static analysis tools via a systematic literature search (\cref{section:literature-search}).
Second, we obtained static and dynamic call graphs for 1000 recent
apps.  To do this, we slightly modified each static analysis tool to extract the call graph it builds.
Then, we ran each static analysis tool on 1000 recent apps (\cref{sec:static_analysis}).
We also ran a dynamic analysis to determine which methods the apps call at
run time (\cref{sec:dynamic_analysis}).
Third, we examined the differences between the statically- and
dynamically-generated call graphs --- in particular, the
methods invoked at run time that do not appear in static models (\cref{sec:empirical_findings}).

Our study seeks to provide directions to the research community on how
can static analysis of Android apps be improved, particularly with recent
apps utilizing external frameworks.

Our results indicate that call graph construction is slow, even when using high-performance computing hardware.
On average, only 58\% of apps can be statically modeled
(only building the call graph, not applying the analysis that uses the call
graph) within
1 hour
by the static analyzers (the average size of apps in our dataset is 24MB and the call graphs computed in our study have an average of \num{16633} nodes and \num{229901} edges).
Tools putatively using the same call graph construction
algorithms have \emph{very} different call graphs and run times.
\emph{Every} static tool suffers significant unsoundness by missing a substantial number of methods in their static model that are called at run time.
Furthermore, more precise call graph construction
implementations suffer more unsoundness.

Although computing precise call graphs is important, it is problematic when it compromises soundness, especially with recent malware using, e.g., implicit call mechanisms.
The Scylla Android malware illustrates how sophisticated threats operate~\cite{scylla}.
It uses the Android framework's JobScheduler to activate payloads only under specific conditions. 
It evades detection by dynamic
analyzers when conditions aren't met. It evades detection by static analyzers that overlook JobScheduler's implicit mechanisms. 
Consequently, such malware can infiltrate platforms like Google Play.
We have investigated the root causes for this unsoundness.
In general, static analyzers suffer from lack of
understanding of the Android framework and external frameworks, leading to missing many implicit calls that, in turn, trigger many methods left unanalyzed statically.

To better illustrate this problem, we ran the tools in our study against the Scylla malware. 
Only 4 of the 13 tools could
statically reach the trigger point of the malicious code using the call
graph.  This  suggests that most current
tools are insufficiently sound to
  handle real-world threats.
This poses a significant risk for end-users, who may
unknowingly use malware that enters
  app markets such as Google Play because the malware is not detected by these tools. Nearly every week, new malware is detected in official app markets.

The main contributions of this paper are as follows:
\begin{itemize}
    \item We conducted a large-scale empirical study on static analysis soundness by comparing call graphs yielded by 13 static analyzers over 1000 Android apps against call graphs yielded by dynamic analysis.
    \item While the problem of unsoundness is known in the literature, we are the first to quantify the extent of the problem in Android app static analysis. In particular, we show that, in the best case, at least 40\% of the methods called at runtime are overlooked by static analyzers.
    \item We show that static analyzers are unsound because they are not properly modeling entry point methods, i.e.,  root nodes in the dynamically generated call graph.
\end{itemize}

\textbf{Findings' Implications:} Previous papers proposing new static analysis techniques (e.g., addressing inter-component communication, reflection, or dynamic loading) often have a formulaic mention of soundness in the threats to validity section, but do not acknowledge the \emph{massive} scale of the problem or the fact that improving soundness is more relevant in practice than improving precision.
Our research findings suggest that:
\dcircle{1} researchers should pause
their work on new analyses and on call graph precision until they have
solved problems of call graph soundness; and
\dcircle{2} library methods are the most often missed by static analysis, so
handling libraries is probably the most important avenue for future work.

As a result, our study sets the stage for more sound static analysis of Android apps, which can lead to safer applications and, ultimately, protect end-users more effectively.

\noindent
\textbf{Artifacts.}
We release all of our artifacts:
\begin{center}
\url{https://github.com/JordanSamhi/Call-Graph-Soundness-in-Android-Static-Analysis}
\end{center}
\section{Motivation}
\label{sec:motivation}

A static analysis uses call graphs to represent the calling relationships between methods.
Most static analyses aim to be sound.  When the analysis is unable to
  resolve a program behavior such as aliasing, method dispatch, reflection,
  etc., a sound analysis must over-approximate the program's possible
  run-time behaviors.  Other challenges for call graph construction in
  Android apps include inter-com\-po\-nent
communication (ICC), callbacks, calls into and out of native code, asynchronous tasks, GUI-related events, etc., for which the target of method calls cannot be inferred statically.
Unsoundness occurs when the estimate of the targets
of a method call omits a target that is called on some execution of
  the call site~\cite{jordan_samhi_thesis_2023}.

Listing~\ref{code:motivation} shows how unsoundness occurs with an implicit call.
An implicit call is a method
call to a method executed at run time without a call to the method in the app's code.
Implicit calls, as established by our research (cf. Section~\ref{sec:cause_unsoundness}), pose a significant obstacle to call graph construction in static analysis.
Omitting them makes a static analyzer
  unaware of certain interactions and dependencies between different parts of the program.

Listing~\ref{code:motivation} contains: 
\dcircle{1} a class called \<MyTask> (line 1) extending the Android framework \<AsyncTask> class; and 
\dcircle{2} an Activity called \<MainActivity> (line 11), the first
Activity launched when the app is run, its \<onCreate()> method is executed (lines 13--16).
On line 14, an object \<myTask> of type \<MyTask> is created.
Then, on line 15, the \<execute()> method is called on the \<myTask> object.
This call will \emph{not trigger} any method called \textit{execute()} in the app's code.
Rather, a mechanism in the Android framework implicitly triggers the execution of method \doInBackground in the app code (lines 3--5).
Furthermore, method \onPostExecute will also be executed after method
\doInBackground has finished, though the app source code has no call to this method.

\begin{listing}
\begin{minted}{java}
|\makebox[0pt][r]{\dcircle{1}\ \strut}|public class MyTask extends AsyncTask<Void, Void, String> {
    @Override
    protected String doInBackground(Void... params) {
        return "Background task completed";
    }
    
    @Override
    protected void onPostExecute(String result) {
        textView.setText(result); // Called after doInBackground
    }
}

|\makebox[0pt][r]{\dcircle{2}\ \strut}|public class MainActivity extends AppCompatActivity {
    @Override
    protected void onCreate(Bundle savedInstanceState) {
        MyTask myTask = new MyTask();
        myTask.execute(); // implicitly calls doInBackground
    }
}
\end{minted}
      \caption{Example: how implicit calls can be triggered in apps.}
    \label{code:motivation}
\end{listing}

Implicit calls present a significant challenge for static analysis. 
They require additional information to model and analyze an app's behavior.
However, it raises the question of how analysts can be expected to possess this knowledge. 
The Android framework, third-party frameworks, and libraries lack
comprehensive documentation regarding the existence and usage of
implicit mechanisms.
This work measures the amount of such methods, among others, missed by existing state-of-the-art static analyzers.
\section{Empirical Study Setup}
\label{sec:empirical_setup}

Our experimental methodology compares static models with methods invoked at run time. 
Dynamic analysis reports actual calls, allowing us to identify unsoundness
in static models: a method that has been called at run time \emph{must}
appear in the static call graph.
Current state-of-the-art static analyzers overlook many methods, showing that statically representing Android apps is not trivial and is still an open problem.
Our goal is to discern why such discrepancies occur and identify the reasons contributing to this problem. 
To this end, our study aims to answer the following research questions:

\begin{description}
    \item[RQ1:] How do static analyzers' call graphs compare to
      each other? 
    \item[RQ2:] How (un)sound are static analyzers?  That is, what proportion of methods executed at run time
      is missed by static analyzers?
    \item[RQ3:] What are the root causes of unsoundness in static analyzers?
    \item[RQ4:] To what extent would approaches collecting callbacks from the Android framework systematically improve the models of static analyzers?
\end{description}

The remainder of this section describes the setup of our study.
Our approach consists of three phases:
\dcircle{1} building call graphs of Android apps using several
state-of-the-art static analyzers (\cref{sec:static_analysis});
\dcircle{2} executing the same apps to collect the methods called at run time (\cref{sec:dynamic_analysis}); and
\dcircle{3} comparing and studying both sets of methods (\cref{sec:empirical_findings}).

\subsection{Dataset}
\label{sec:dataset}

We randomly collected a representative sample of \num{1000} recent real-world apps from the \num{24000000}+ APKs in the AndroZoo repository~\cite{10.1145/2901739.2903508} (with a confidence level of 99\% and a confidence interval of $\pm$ 5\%), all of which were collected in 2023 by AndroZoo from app markets.
The average size of apps in our dataset is \num{24}MB, the median is \num{15}MB, and the standard deviation is \num{24}MB. 
These 1000 apps contain an
    average of \num{132687} methods, an average of \num{108973} non-library methods, and the medians are, respectively,
    \num{110378} and \num{96571}.
Our dataset is available in our project's
artifacts,
along with our experimental framework.

\subsection{Tools}
\label{section:literature-search}

Our selection criteria for tools included:
\dcircle{1} tools relying on call graphs; and
\dcircle{2} open-source (which allows us to slightly modify them for extracting their call graphs).

To gather tools, we relied on a frequently used strategy in systematic
literature reviews (SLRs): identifying relevant keywords to help us
identify a large number of potential papers from well-known databases.
We searched for  "android" AND "static" AND ("callgraph" OR "call graph" OR
"call-graph" OR "model" OR "implicit" OR "callback" OR "component"). 
We searched in three databases: IEEE Xplore, ACM DL, and Science Direct.
This search yielded 423 distinct papers, which describe 61 static analysis tools using a call graph. 
Among these tools, 21/61 (34\%) match our criteria (i.e., use a call graph, and are open-source).

\begin{table}[h!]
\begin{adjustbox}{width=.9\columnwidth,center}
        \centering
            \begin{tabular}{|lc|lc|lc|}
                \hline
                Tool & Runnable & Tool & Runnable & Tool & Runnable \\
                \hline
                ACID~\cite{10.1145/3510454.3516854}         & \ding{51} & DroiDel~\cite{10.1145/2771284.2771288}      &           &  IccTA~\cite{10.5555/2818754.2818791} & \ding{51}  \\
                AppoScopy~\cite{10.1145/2635868.2635869}    &           & DroidRA~\cite{10.1145/2931037.2931044}      & \ding{51} &   Jicer~\cite{9610738}           &    \ding{51}        \\
                ArpDroid~\cite{10.1145/3197231.3197255}     & \ding{51} & DroidSafe~\cite{gordon2015information}      & \ding{51} & MaMaDroid~\cite{10.1145/3313391} & \ding{51}   \\
                BackDroid~\cite{9505167}                    & \ding{51} & ELEGANT~\cite{8719450}                      &           &    NatiDroid~\cite{10.1145/3540250.3549142}                  & \ding{51}  \\ 
                BackStage~\cite{10.1145/3197231.3197232}    &           & FlowDroid~\cite{10.1145/2666356.2594299}  & \ding{51} & NaDroid~\cite{10.1145/3168829} & \\
                DidFAIL~\cite{klieber2014android}           &           & Gator~\cite{7194564}                      & \ding{51} & RAICC~\cite{10.1109/ICSE43902.2021.00126} & \ding{51}\\
                Difuzer~\cite{10.1145/3510003.3510135}      & \ding{51} & HybriDroid~\cite{10.1145/2970276.2970368} &           & SootFX~\cite{9610670} & \ding{51}\\
                \hline
            \end{tabular}
    \end{adjustbox}
    \caption{Open-source static analysis tools that use a call graph.
      ``Runnable'' means we could adapt, build, and run the tool.}
    \label{tab:tools}
\end{table}

Table~\ref{tab:tools} shows the list of tools that match our criteria.
Among these tools, we were unable to build 6 of them (i.e., Hybridroid,
ELEGANT, Backstage, DroidDEL, DIDFail, and AppoScopy) and unable to run one of them
(i.e., NaDroid).
We have contacted the authors of these 7 tools.
We have received three answers, among which two
to help us build the tools (DidFail and ELEGANT), but eventually we were not able to build 
them, and in the third answer regarding Backstage, one of the authors said that they will try to contact the developer, we have not received any news so far.

We ran 14 static analysis tools on all the apps, but we dropped DroidSafe from our experiments as it could only analyze 6 out of the \num{1000} apps.
Therefore, this paper's results contain \numtools tools.

Note that while each considered tool computes a call graph, 
each tool is built differently and may have been designed for different goals. 
We briefly describe the main goal of each tool in Table~\ref{tab:toolsDescription}. 
The supplementary material gives the configuration of each tool as run in our experiments.

\begin{table}[h!]
\begin{adjustbox}{width=\columnwidth,center}
        \centering
            \begin{tabular}{lll}
(1) & FlowDroid& detects data leaks in Android apps \\ 
(2) & IccTA &detects potential data leaks in apps with an ICC sensitivity\\
(3) & RAICC & extends FlowDroid with additional ICC methods \\
(4) & DroidRA & extends FlowDroid to resolve reflective calls and improve call graphs \\
(5) & NatiDroid& performs cross-language static analyses of both  bytecode and native code \\
(6) & MaMaDroid & detects malware based on app behavior \\
(7) & BackDroid & on-the-fly bytecode search to improve inter-procedural analysis \\
(8) & SootFX & extracts features for machine learning \\
(9) & ACID & detects API compatibility issues \\
(10) &  Gator & performs callback-sensitive static analysis\\
(11) & Jicer &is a bytecode slicer\\
(12) & ArpDroid& detects and repairs incompatible uses of the runtime permission \\
(13) & Difuzer & detects hidden sensitive information in apps (via data flow analysis) \\
            \end{tabular}
    \end{adjustbox}
    \caption{Description of the tools.}
    \label{tab:toolsDescription}
\end{table}

\subsection{Running Static Analysis}
\label{sec:static_analysis}

For each static analysis tool
that we have considered, we modified them slightly to extract relevant data, hence our modifications have no impact on the analyses.
Note that several tools, such as DroidRA, modify the original call graph used.
Hence, before extracting the call graphs, we let the tools exercise the call graphs so that we extract the call graphs they use for their analyses.

We then ran the modified tool (with different call graph construction
algorithms when possible; see \cref{fig:num_apps_succesfully_analyzed,table:results_statistic_common_apps} for all 25 configurations) with a 1-hour timeout and with the default configuration described by the developers.
Prior to using the 1-hour timeout, we performed the same experiment with a 10-minute timeout.
The findings remained unchanged.
Moreover, with a 1-hour timeout, the number of successfully analyzed apps did not increase.
Thus, increasing the timeout is unlikely to 
\dcircle{1} change the findings of this paper; and
\dcircle{2} significantly increase the number of successfully analyzed apps.
The interested reader can check our artifacts for all the results.
Note that, when the timeout is reached, there is no call-graph, analyzers do not yield a partial call graph.

The output of this process is, for each app and each call graph
construction algorithm, six sets of methods and two sets of edges:

\label{sec:variable-definitions}
\begin{enumerate}
    \item $\|SM|$: the set of all methods in the app (to obtain this information, we have iterated over all classes and counted all methods present in these classes)\footnote{We compute these sets using the Jimple intermediate representation.}
    \item $\|SM|_{\|cg|}$: the set of all methods in the call graph
    \item $\|SM|_{\neg \|cg|} = \|SM| - \|SM|_{\|cg|}$: the set of all methods that do not appear in the call graph
    \item $\|SM|^{\neg l}$: set of all non-library methods
    \item $\|SM|_{\|cg|}^{\neg l}$: set of all non-library methods in the call graph 
    \item $\|SM|_{\neg \|cg|}^{\neg l} = \|SM|^{\neg l} - \|SM|_{\|cg|}^{\neg l}$: the set of all methods that are neither in the call graph nor classified as libraries
    \item $\|SE|$: the set of all edges in the call graph
    \item $\|SE|^{\neg l}$ the set of all edges in the call graph whose targets are non-library methods.
\end{enumerate}

To determine whether a method is a library or not, we relied on the list of Android libraries given in~\cite{androlibzoo}.

\subsection{Running Dynamic Analysis}
\label{sec:dynamic_analysis}

\begin{listing}
    \centering
\begin{minted}{java}
public void m() {
+ Log.d("MY_LOGGER", "<Class: void m()>");
  // method body
+ Log.d("MY_LOGGER", "<Class: void m()>--><Cat: void <init>()>");
  Cat c = new Cat();
+ Log.d("MY_LOGGER", "<Class: void m()>--><Cat: void walk()>");
  c.walk();
}
\end{minted}
    \caption{Instrumentation for logging methods called at run time.}
    \label{code:instrumentation}
\end{listing}

We built a dynamic call graph analysis.
Its main component is an instrumentation tool that inserts a log statement at:
\dcircle{1} the beginning of each method in the app; and
\dcircle{2} each method call \emph{in the app}.
The dynamic call graph analysis uses the log to construct three sets per app:
$\|DM|$ a set of methods called;
$\|DM|^{\neg l}$ a set of non-library methods called; and
$\|DE|$ a set of dynamically collected edges.
Note than only the code inside apps is instrumented, the Android framework is not.

As an example, consider the code in Listing~\ref{code:instrumentation}.  On
line 2, a simple log statement is added with the name of the method (in
Jimple format) that is being executed.
Lines 4 and 6 demonstrate how our instrumentation tool would insert log statements to record that the current method is invoking the method called within it.
Our implementation is more sophisticated; for example, it correctly
handles the case when the call is in a subexpression that might or might
not get executed.
Our implementation does not record the actual targets of reflective calls.
At a call that is executed via dynamic dispatch, the target is recorded as
it appears in the bytecode, rather than all the method implementations among
which dispatch might choose.
Missing calls in the dynamic call graph mean that this paper may
under-report the unsoundness of static analysis tools.

After instrumentation, we signed each app and installed it on a headless
Android emulator based on Google's android-33 system image (x86\_64).

Subsequently, we exercised each app for 5 min with Monkey~\cite{monkey}, generating random inputs.
We used Monkey because of empirical evidence~\cite{10.1145/3460319.3464828,9000056} that, despite the existence of more complex approaches to augment code coverage, Monkey still achieves the best coverage performance in practice.
Also, a recent study has shown that after 5 min, the proportion of code covered using Monkey reaches a ``plateau''~\cite{9000056}.

\label{sec:dynamic-analysis-low-coverage}

The dynamic analysis observed \num{310595043} method calls (including implicit invocations in the apps)
to \num{1082265}
unique methods across the \num{1000} apps of our dataset\footnote{For
instance, a method m(), i.e., a given method implementation, can be called $n$ times, but is only counted once in the set of unique methods.}.
The average code coverage (at the method level) for our dynamic analysis over all 1000 apps is 8\% and the median is 4\%.
We acknowledge that the code coverage is low.
However, we remind the reader that our goal is not to reach high code coverage, and that a low code coverage will actually reinforce our findings.
Indeed, if the number of executed methods is low, and if static analyzers
miss a high proportion of these methods, then the problem
of unsoundness that this paper reports is a
lower bound of the actual problem.

\begin{figure}[ht]
    \centering
    \includegraphics[scale=.26]{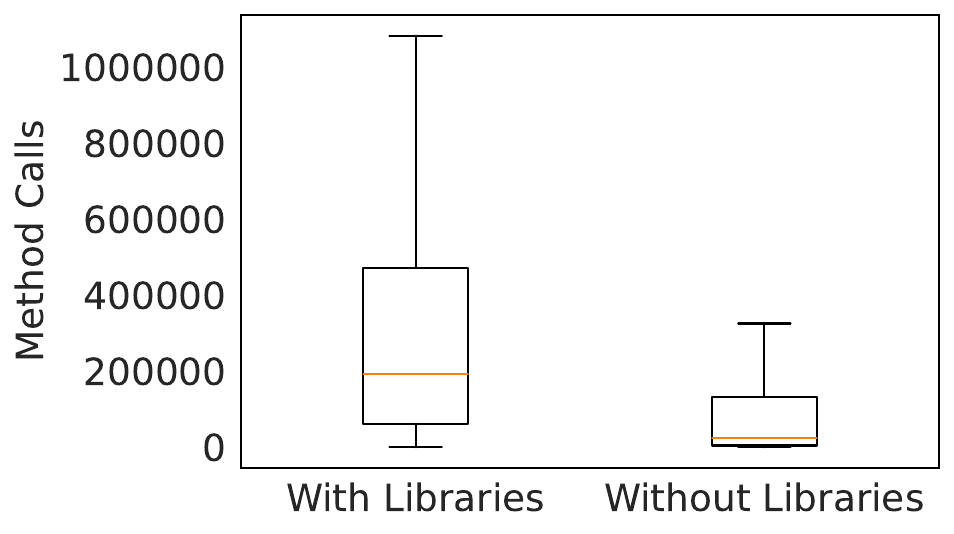}
    \includegraphics[scale=.26]{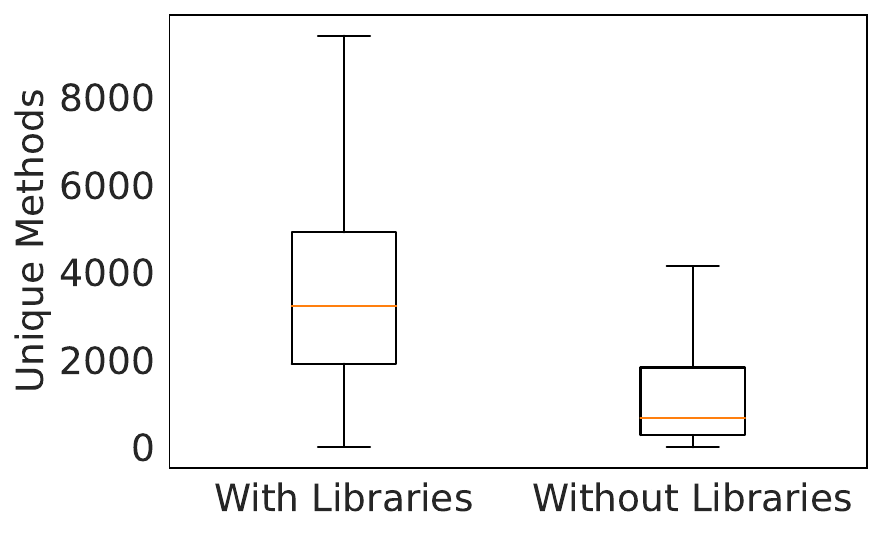}
    \caption{The number of methods called at run time.}
    \label{fig:distributions_methods_per_app}
\end{figure}

Figure~\ref{fig:distributions_methods_per_app} shows the distributions of
the number of method calls and unique methods called during the dynamic analysis
with and without library methods.
Table~\ref{tab:mean_median_methods} shows the mean and median numbers of
method calls and unique methods collected from the dynamic
analysis.
Results indicate that, on average, there are more than twice as many library calls in Android apps during execution as non-library calls.

\begin{table}[ht]
\centering
\begin{adjustbox}{width=.7\columnwidth,center}
\begin{tabular}{|l|c|c|}
\hline
 & \textbf{Mean} & \textbf{Median} \\ 
\hline
\textbf{Method Calls ($\|DM|$)} & 310 595 & 191 862 \\ 
\hline
\textbf{Unique Methods} & 3 885 & 3 236 \\
\hline
\textbf{Non Library Method Calls ($\|DM|^{\neg l}$)} & 136 374 & 24 310 \\
\hline
\textbf{Non Library Unique Methods} & 1 355 & 678 \\
\hline
\end{tabular}
\end{adjustbox}
\caption{Mean and Median of the number of method calls and the number of unique methods with and without libraries}
\label{tab:mean_median_methods}
\end{table}
\section{Empirical Findings}
\label{sec:empirical_findings}

\subsection{RQ1: Comparison of Static Analyzers Model}
\label{sec:rq1}

\begin{figure}[h]
    \centering
    \begin{adjustbox}{width=.9\columnwidth,center}
    \includegraphics{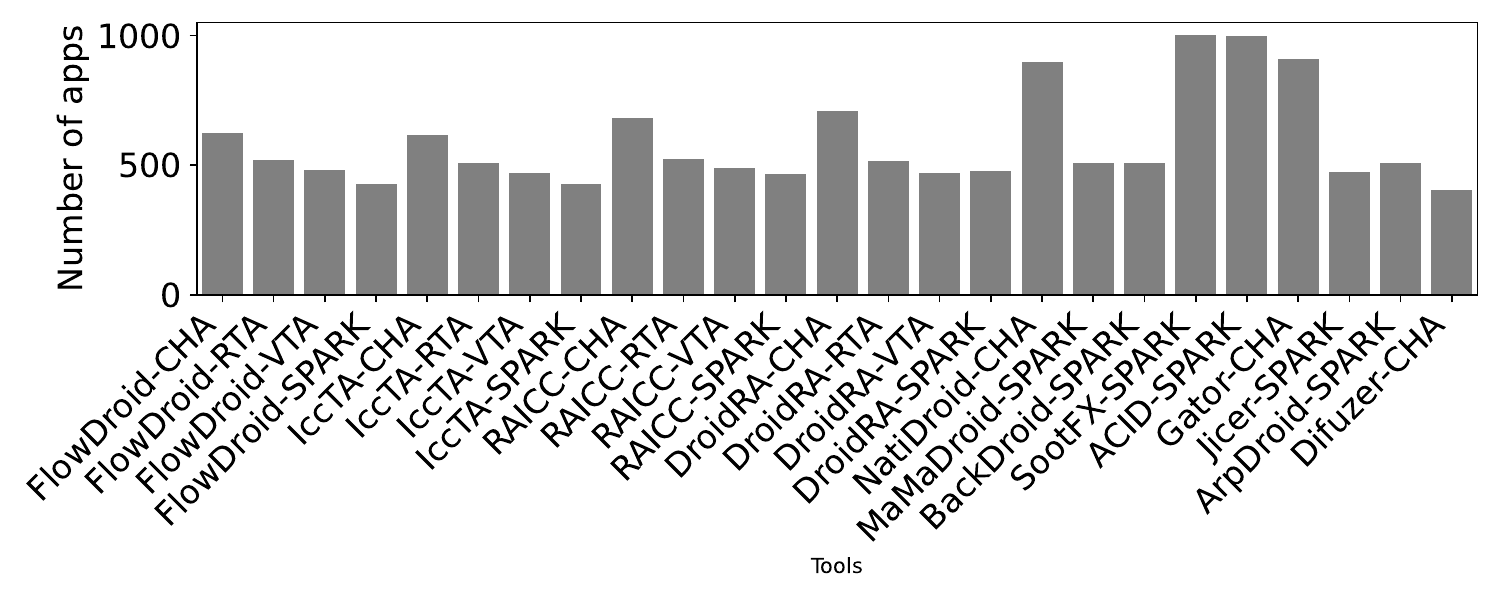}
    \end{adjustbox}
    \caption{Number of apps successfully analyzed per tool}
    \label{fig:num_apps_succesfully_analyzed}
\end{figure}

Figure~\ref{fig:num_apps_succesfully_analyzed} reports the number of apps successfully analyzed by each tool.
For instance, FlowDroid extracted 624 call graphs
using the CHA algorithm and 429 call graphs using the SPARK algorithm.
For the remaining apps, i.e., $1000 - 624 = 376$ for FlowDroid-CHA, FlowDroid crashed 7 times and reached the timeout 369 times.
For SPARK, FlowDroid reached the timeout for 566 apps, and crashed for 5 apps (the detailed results are available in our artifacts).
Overall, most of the considered static analysis tools can only successfully analyze about half of the 1000
  apps.  This is a threat to the validity of previous work that used these
  tools, which may not have been run on a representative sample of apps.
  The most robust tools are NatiDroid, SootFX, ACID, and Gator.

Only
126 apps were successfully analyzed by every tool.
The remainder of this paper focuses on those apps, in order to
permit a fair comparison among the tools.  The supplementary
  material provides the results for all apps.
These 126 apps are
  probably among the simplest of the 1000, since they did not trigger bugs
  or limitations in the static analysis tools.  Therefore, the actual
  unsoundness of static analysis tools in practice is probably worse than
  reported by this paper.

\Cref{table:results_statistic_common_apps} provides data about the
  apps together with their libraries (the "With libraries"
columns), and also data about only the app code (the ``Without
  libraries'' columns).  Both sets columns are
  data from the same analysis run, but the ``Without
  libraries'' columns only count the part of the computed call graph that
  comes from developer code.
Each $|\|SM||$ is the number of methods that the analysis tool discovered
  in the app.
The ``\%M in CG'' columns report how many of those methods appear in
  the call graph: $|\|SM|_{\|cg|}| / |\|SM||$ (see
  \cref{sec:static_analysis} for definitions).
provide statistics about the methods identified by the tools within the successfully analyzed apps.

\begin{table}
    \centering
    \caption{Methods gathered statically in apps, for 126 apps. (CG = call graph, M. = Methods,
     Avg. = Average). The denominator for the \% columns is the number of methods in apps.}
    \begin{adjustbox}{width=\columnwidth,center}
        \begin{tabular}{|c|c||c|c|c||c|c|c|}
            \hline
            &  &
            \multicolumn{3}{c||}{\textbf{With libraries}} & 
            \multicolumn{3}{c|}{\textbf{Without libraries}} \\ \cline{3-8}
            &  &
                \multicolumn{1}{c|}{\textbf{Avg.}} &
                \multicolumn{1}{c|}{\textbf{\% M.}} & 
                \multicolumn{1}{c||}{\textbf{Avg.}} &
                \multicolumn{1}{c|}{\textbf{Avg.}} &
                \multicolumn{1}{c|}{\textbf{\% M.}} & 
                \multicolumn{1}{c|}{\textbf{Avg.}}
                \\  
            & & \textbf{$|\|SM||$}  &  \textbf{in CG} & \textbf{$|\|SE||$}
            & \textbf{$|\|SM|^{\neg l}|$}  &   \textbf{in CG} &\textbf{$|\|SE|^{\neg l}|$}  \\ \hline
                    \multirow{4}{*}{FlowDroid} & CHA
                    & \num{71051}
					& \num{38}\%
					& \num{399975}
					& \num{6651}
					& \num{66}\%
					& \num{48218}
                     \\ \cline{2-2} \cline{3-8}
                     & RTA
                    & \num{71046}
					& \num{24}\%
					& \num{227493}
					& \num{6651}
					& \num{52}\%
					& \num{33802}
                     \\ \cline{2-2} \cline{3-8}
                     & VTA
                    & \num{71045}
					& \num{18}\%
					& \num{109519}
					& \num{6651}
					& \num{42}\%
					& \num{16788}
                     \\ \cline{2-2} \cline{3-8}
                     & SPARK
                    & \num{71031}
					& \num{5}\%
					& \num{15250}
					& \num{6649}
					& \num{12}\%
					& \num{2391}
                     \\ \hline \hline
                    \multirow{4}{*}{IccTA} & CHA
                    & \num{71051}
					& \num{38}\%
					& \num{399981}
					& \num{6651}
					& \num{66}\%
					& \num{48220}
                     \\ \cline{2-2} \cline{3-8}
                     & RTA
                    & \num{71046}
					& \num{24}\%
					& \num{227541}
					& \num{6651}
					& \num{52}\%
					& \num{33746}
                     \\ \cline{2-2} \cline{3-8}
                     & VTA
                    & \num{71045}
					& \num{18}\%
					& \num{109023}
					& \num{6651}
					& \num{41}\%
					& \num{16703}
                     \\ \cline{2-2} \cline{3-8}
                     & SPARK
                    & \num{71031}
					& \num{5}\%
					& \num{15249}
					& \num{6649}
					& \num{12}\%
					& \num{2391}
                     \\ \hline \hline
                    \multirow{4}{*}{RAICC} & CHA
                    & \num{71051}
					& \num{38}\%
					& \num{397791}
					& \num{6651}
					& \num{66}\%
					& \num{47894}
                     \\ \cline{2-2} \cline{3-8}
                     & RTA
                    & \num{71046}
					& \num{24}\%
					& \num{224574}
					& \num{6651}
					& \num{52}\%
					& \num{33271}
                     \\ \cline{2-2} \cline{3-8}
                     & VTA
                    & \num{71045}
					& \num{19}\%
					& \num{111151}
					& \num{6651}
					& \num{41}\%
					& \num{16605}
                     \\ \cline{2-2} \cline{3-8}
                     & SPARK
                    & \num{71031}
					& \num{6}\%
					& \num{16264}
					& \num{6650}
					& \num{12}\%
					& \num{2434}
                     \\ \hline \hline
                    \multirow{4}{*}{DroidRA} & CHA
                    & \num{71053}
					& \num{38}\%
					& \num{397872}
					& \num{6652}
					& \num{66}\%
					& \num{47903}
                     \\ \cline{2-2} \cline{3-8}
                     & RTA
                    & \num{71048}
					& \num{24}\%
					& \num{224992}
					& \num{6652}
					& \num{52}\%
					& \num{33452}
                     \\ \cline{2-2} \cline{3-8}
                     & VTA
                    & \num{71047}
					& \num{19}\%
					& \num{111188}
					& \num{6652}
					& \num{42}\%
					& \num{16749}
                     \\ \cline{2-2} \cline{3-8}
                     & SPARK
                    & \num{71033}
					& \num{6}\%
					& \num{16437}
					& \num{6650}
					& \num{12}\%
					& \num{2491}
                     \\ \hline \hline
                    \multirow{1}{*}{NatiDroid} & CHA
                    & \num{61758}
					& \num{81}\%
					& \num{469025}
					& \num{4837}
					& \num{88}\%
					& \num{40398}
                     \\ \hline \hline
                    \multirow{1}{*}{MaMaDroid} & SPARK
                    & \num{60500}
					& \num{5}\%
					& \num{12592}
					& \num{4791}
					& \num{14}\%
					& \num{2007}
                     \\ \hline \hline
                    \multirow{1}{*}{BackDroid} & SPARK
                    & \num{60500}
					& \num{5}\%
					& \num{12592}
					& \num{4791}
					& \num{14}\%
					& \num{2007}
                     \\ \hline \hline
                    \multirow{1}{*}{SootFX} & SPARK
                    & \num{61707}
					& \num{0}\%
					& \num{101}
					& \num{4798}
					& \num{1}\%
					& \num{9}
                     \\ \hline \hline
                    \multirow{1}{*}{ACID} & SPARK
                    & \num{61707}
					& \num{8}\%
					& \num{54169}
					& \num{4798}
					& \num{48}\%
					& \num{4124}
                     \\ \hline \hline
                    \multirow{1}{*}{Gator} & CHA
                    & \num{110824}
					& \num{73}\%
					& \num{1920412}
					& \num{31342}
					& \num{90}\%
					& \num{655813}
                     \\ \hline \hline
                    \multirow{1}{*}{Jicer} & SPARK
                    & \num{71144}
					& \num{6}\%
					& \num{15763}
					& \num{6651}
					& \num{11}\%
					& \num{2302}
                     \\ \hline \hline
                    \multirow{1}{*}{ArpDroid} & SPARK
                    & \num{60500}
					& \num{5}\%
					& \num{12593}
					& \num{4791}
					& \num{14}\%
					& \num{2007}
                     \\ \hline \hline
                    \multirow{1}{*}{Difuzer} & CHA
                    & \num{60567}
					& \num{34}\%
					& \num{245987}
					& \num{4809}
					& \num{65}\%
					& \num{31060}
                     \\ \hline
        \end{tabular}
    \end{adjustbox}
    \label{table:results_statistic_common_apps}
\end{table}

Exhaustively comparing the tools is outside the scope of this paper,
because our objective is to quantify unsoundness by using dynamic analysis
as a ground truth. Nonetheless, we note a few observations from the data.

\textbf{Different tools find different methods $|\|SM||$ in an app.}
One set of tools (ACID, ArpDroid, BackDroid, Difuzer, MaMaDroid,
  NatiDroid, and SootFX) find about\num{61000}
methods per app; another set of tools (DroidRA, FlowDroid, IccTA, Jicer,
and RAICC) finds about \num{71000}
per app, and Gator finds \num{111000}.
We have investigated and could find the following explanation for these differences.
Firstly, we confirm that none of the following tools: ACID, ArpDroid, BackDroid, Difuzer, MaMaDroid,
  NatiDroid, and SootFX consider all dex files in apps for the static analysis, they only consider the main ``classes.dex'' file and therefore miss many methods if additional ``.dex'' files are present in apps, which explain the low number of methods discovered statically compared to the rest.
Secondly, we confirm that the following tools: DroidRA, FlowDroid, IccTA, Jicer,
and RAICC consider all ``.dex'' files in apps which explains why they are able to find about \num{10000} additional methods compared to the first set of tools.
For the last tool, i.e., Gator, we confirm that it also consider all ``.dex'' files in apps but could not find any additional hint about why it finds \num{40000} additional methods in apps.

\textbf{IccTA, RAICC, and DroidRA add few edges to the call graph.}
These tools
are built upon FlowDroid and are designed to add edges to
its call graph; in other words, they are intended to correct
  unsoundness in FlowDroid.
Corresponding rows in \cref{table:results_statistic_common_apps} are
  little different --- always well under 10\% and usually closer to 0\%
  different.  As shown later in this paper, they do not address the most
  important sources of unsoundness in FlowDroid.
In some cases, RAICC-CHA and DroidRA-CHA have \emph{fewer} edges than
FlowDroid-CHA, even though those algorithms are designed to \emph{add}
edges.  We do not have an explanation for this behavior, though
nondeterminism may play a part~\cite{Soles2023}.

\textbf{More precise algorithms succeed in pruning the call graph.}
SPARK is more precise than VTA, VTA than RTA, and RTA
than CHA.  \Cref{table:results_statistic_common_apps} shows that the
more precise algorithms lead to graphs containing fewer methods and fewer edges.
This aligns with the fact that enhancing precision results in reducing over-approximation.

\textbf{Static analysis considers large portions of apps to be dead code.}
In the ``With libraries'' columns, most code is
dead code (i.e., methods not in call graph), likely because
no app exercises all parts of a library that it depends on.
More surprising are the ``Without libraries'' columns, where FlowDroid
considers apps to contain 1/3 to 7/8 dead code (depending on the call graph
construction algorithm).  This likely reflects unsoundness (many methods
  are present but are not modeled statically in the call graph), since it seems
unlikely that developers would ship 8 times as much code as an app needs,
most of it useless.  NatiDroid and Gator more plausibly claim that 88--90\%
of the app code may be executed.

\textbf{Implementations of the same call graph construction algorithm
    differ.}
Different implementations of an algorithm may yield slightly different
results due to different modeling and implementation choices.  The
implementations of CHA in FlowDroid and Difuzer retain very similar
percentages of methods.  However, the CHA implementations in NatiDroid and
Gator retain dramatically more methods.
In the ``Without libraries'' columns, all SPARK implementations retain
11--14\% of methods --- except SootFX which retains 1\% and ACID which
retains 48\%.
We investigated SootFX and found that its problem is its set of entry
  points:  SootFX uses only \<Threads> as an entry point, which misses most
  of the application.  The set of entry points is at least as important as
  the call graph construction algorithm.

\highlight{
\textbf{RQ1 answer:}
Our comparison of the static analyzers shows that:
\dcircle{1} only a small proportion, i.e., 58\% on average, of apps can be analyzed by static analysis tools in a 1-hour timeframe;
\dcircle{2} static analysis approaches supposed to improve call graphs' soundness show little variation in the number of methods and edges in their call graphs; and
\dcircle{3} although static analysis tools share similar call graph construction algorithms, they exhibit different (sometimes significantly divergent) call graphs.
}

\subsection{RQ2: Unsoundness: Methods Missed by Static Analyzers}
\label{sec:rq2}

\begin{figure}[ht]
  \centering
    \includegraphics[width=.9\columnwidth]{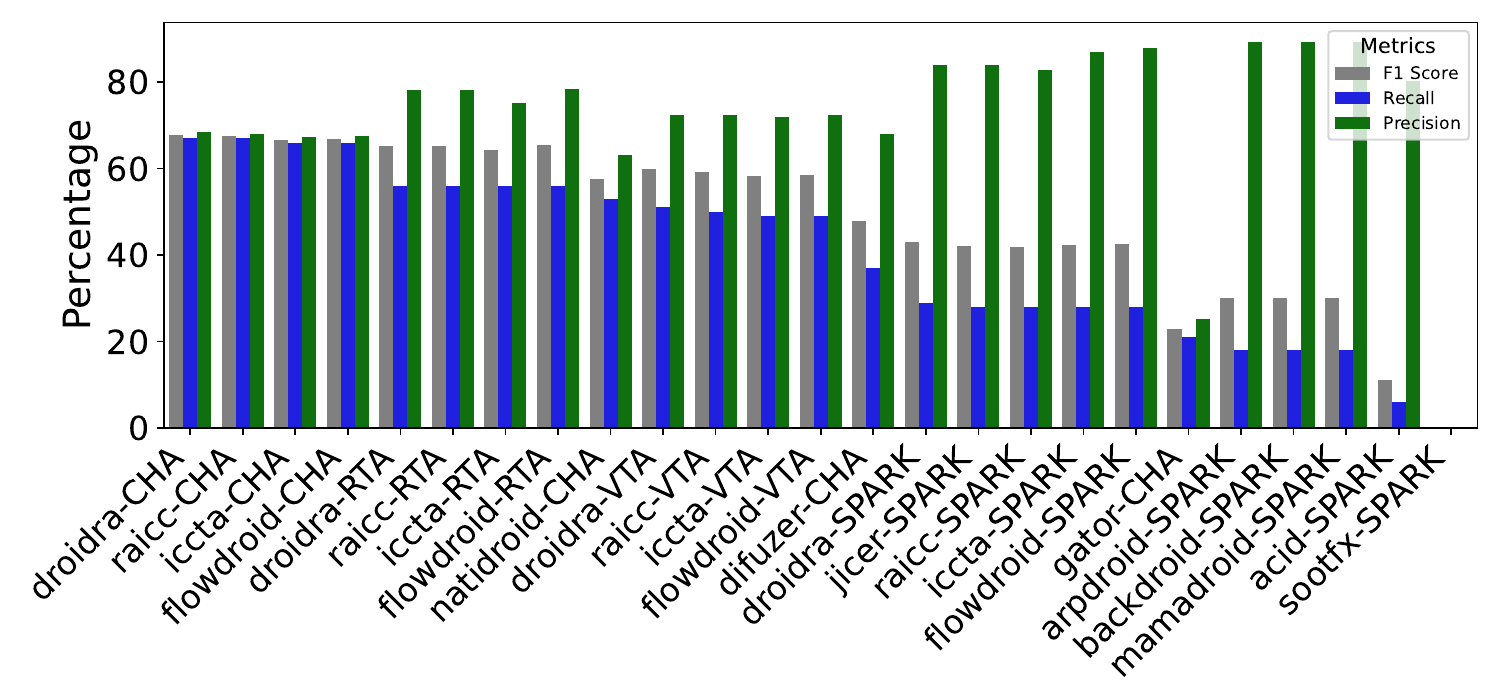}
  \caption{Comparison of recall, precision, and f$_{1}$ score of all configurations of tools and call graph construction algorithms.}
\label{fig:tool_cg_proportion_methods_missed_dynamic_common_apps}
\end{figure}

\Cref{fig:tool_cg_proportion_methods_missed_dynamic_common_apps} shows the precision, recall, and f$_1$\ score of the methods in static call graphs compared to \emph{methods in dynamic call graphs} (since the dynamic analysis does not yield complete call graphs given that coverage is limited, we report precision with respect to the call graphs collected dynamically) for each app and each configuration, i.e., tools and call graph construction algorithms.
For a fair head-to-head comparison,
\cref{fig:tool_cg_proportion_methods_missed_dynamic_common_apps} reports
results for the 126 apps that were analyzed by all tools.
The supplementary material provides the results for all apps successfully analyzed by each tool.

The goal of a more precise call graph construction algorithm, such as
  SPARK, is to soundly improve the precision of the call graph.  That is, a
  more precise algorithm should remove infeasible nodes and edges from the
  call graph \emph{without} removing feasible ones.
  \Cref{fig:tool_cg_proportion_methods_missed_dynamic_common_apps} shows
  that these algorithms fail at their goal.  While they do improve
  precision, they do so at the cost of increased unsoundness. More
  specifically, in
  \cref{fig:tool_cg_proportion_methods_missed_dynamic_common_apps}, SPARK
  tends to have the greatest unsoundness, CHA tends to have the least, and
  RTA and VTA are in between.  But even the CHA-based tools report
  unacceptable unsoundness with 21\%-67\% recall.  We were not able to determine whether the
  more precise algorithms are fundamentally flawed (as fielded in the given
  tools), or the implementations of the algorithms are defective.  Although
  the more precise algorithms are more complex to implement, we did not
  expect that complexity to lead to systematically more bugs.

One outlier is the CHA-based Gator tool, which has nearly  the highest
level of unsoundness (lowest recall).  After investigation, we found that
Gator relies on a unique two-step technique to build its call graph: first,
it considers all methods within the app as a potential entry point which is
a considerable over-approximation; and second, it refines the set of
potential entry points with many rules based on permissions, the manifest,
etc.  Future tools should not adopt its novel call graph construction
technique.

Our finding shows that (in current
  implementations) the more precise the algorithm, the more unsound: it misses many methods, i.e., apps' code, during analysis.
Most research in call graph construction algorithms has the goal of
  improving precision.  Our research throws doubt on the
desirability of precision-focused call graph construction
algorithms since they are time-consuming to build and often result in a
significant proportion of code being overlooked.  They are
  strictly worse for analyses related to security, among other applications.

We again remind the reader that the methods collected
dynamically were gathered in a mere 5 minutes of random execution. 
Hence, our findings likely underestimate the volume of methods static analyzers miss.

\begin{table}
    \centering
    \caption{Methods that were called at run time but not modeled statically.}
    \label{tab:dynamic_not_static}
    \begin{adjustbox}{width=.9\columnwidth,center}
    \begin{tabular}{|c|c||c|c|c|c|c|}
        \hline
        \multirow{2}{*}{Tool} & \multirow{2}{*}{Algorithm} & \multicolumn{5}{c|}{Missing methods ($\|DM| \setminus \|SM|_{\|cg|}$)} \\ \cline{3-7}
        & & Mean & Median & Min & Max & Total \\ \hline
        \multirow{4}{*}{FlowDroid} 
        & CHA & \num{1294} & \num{685} & \num{14} & \num{14397} & \num{163099} \\ \cline{2-7}
        & RTA & \num{1499} & \num{936} & \num{29} & \num{14399} & \num{188845} \\ \cline{2-7}
        & VTA & \num{1621} & \num{1110} & \num{41} & \num{14399} & \num{204220} \\ \cline{2-7}
        & SPARK & \num{1965} & \num{1216} & \num{42} & \num{14430} & \num{247577} \\ \hline \hline
        \multirow{4}{*}{IccTA} 
        & CHA & \num{1294} & \num{685} & \num{14} & \num{14397} & \num{163094} \\ \cline{2-7}
        & RTA & \num{1498} & \num{938} & \num{29} & \num{14399} & \num{188799} \\ \cline{2-7}
        & VTA & \num{1621} & \num{1110} & \num{41} & \num{14399} & \num{204268} \\ \cline{2-7}
        & SPARK & \num{1965} & \num{1216} & \num{42} & \num{14430} & \num{247577} \\ \hline \hline
        \multirow{4}{*}{RAICC} 
        & CHA & \num{1295} & \num{684} & \num{14} & \num{14397} & \num{163225} \\ \cline{2-7}
        & RTA & \num{1522} & \num{983} & \num{41} & \num{14399} & \num{191785} \\ \cline{2-7}
        & VTA & \num{1614} & \num{1064} & \num{41} & \num{14399} & \num{203374} \\ \cline{2-7}
        & SPARK & \num{1935} & \num{1188} & \num{42} & \num{14430} & \num{243871} \\ \hline \hline
        \multirow{4}{*}{DroidRA} 
        & CHA &  \num{1295} & \num{683} & \num{14} & \num{14397} & \num{163195} \\ \cline{2-7}
        & RTA & \num{1521} & \num{976} & \num{41} & \num{14399} & \num{191606} \\ \cline{2-7}
        & VTA & \num{1612} & \num{1064} & \num{41} & \num{14399} & \num{203174} \\ \cline{2-7}
        & SPARK & \num{1933} & \num{1188} & \num{42} & \num{14430} & \num{243570} \\ \hline \hline
        \multirow{1}{*}{NatiDroid} 
        & CHA & \num{1299} & \num{704} & \num{0} & \num{14382} & \num{163639} \\ \hline \hline
        \multirow{1}{*}{MaMaDroid} 
        & SPARK & \num{2085} & \num{1344} & \num{42} & \num{14430} & \num{262671} \\ \hline \hline
        \multirow{1}{*}{BackDroid} 
        & SPARK & \num{2085} & \num{1344} & \num{42} & \num{14430} & \num{262671} \\ \hline \hline
        \multirow{1}{*}{SootFX} 
        & SPARK & \num{2641} & \num{1828} & \num{84} & \num{15322} & \num{332824} \\ \hline \hline
        \multirow{1}{*}{ACID} 
        & SPARK & \num{2286} & \num{1704} & \num{84} & \num{15322} & \num{288068} \\ \hline \hline
        \multirow{1}{*}{Gator} 
        & CHA & \num{2213} & \num{1462} & \num{6} & \num{14815} & \num{278840} \\ \hline \hline
        \multirow{1}{*}{Jicer} 
        & SPARK & \num{1962} & \num{1194} & \num{42} & \num{14430} & \num{247237} \\ \hline \hline
        \multirow{1}{*}{ArpDroid} 
        & SPARK & \num{2085} & \num{1344} & \num{42} & \num{14430} & \num{262671} \\ \hline \hline
        \multirow{1}{*}{Difuzer} 
        & CHA & \num{1542} & \num{926} & \num{14} & \num{14397} & \num{194256} \\ \hline
\end{tabular}\end{adjustbox}
\end{table}

The ``Total'' column in
\cref{tab:dynamic_not_static} is proportional to missed methods, and
inversely proportional to soundness, in
\cref{fig:tool_cg_proportion_methods_missed_dynamic_common_apps}.
\Cref{tab:dynamic_not_static} breaks the data down more finely.

We can make several observations:
\dcircle{1} although CHA is the highest degree of over-approximation in existing call graph construction algorithms, it still falls short in capturing a large quantity of methods;
\dcircle{2} SPARK, the most precise call-graph construction algorithm for Android apps, falls behind, missing almost twice as many methods on average as CHA does, i.e., SPARK is less sound than CHA (and RTA and VTA); and
\dcircle{3} with SootFX and ACID
at least one app shows up to \num{15322} methods missing, a statistic that could prove critical if the app is, in fact, a malware.

\begin{figure}[ht]
    \centering
    \begin{adjustbox}{width=.9\columnwidth,center}
    \includegraphics{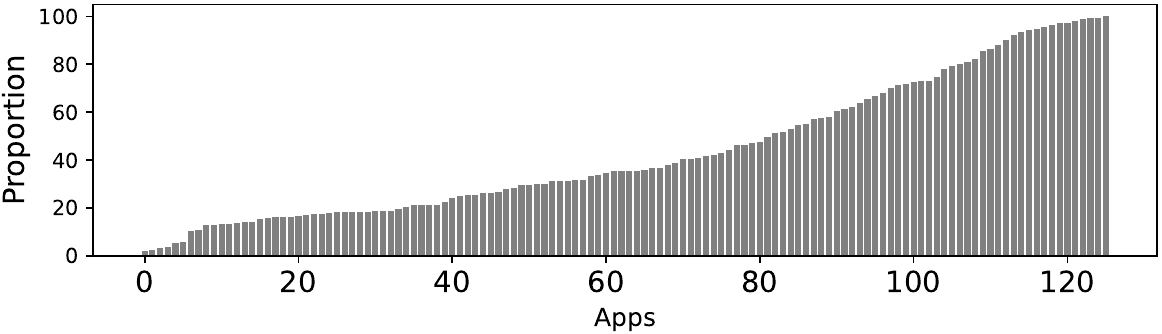}
    \end{adjustbox}
    \caption{Proportion of dynamically-executed methods missed by FlowDroid-CHA.}
    \label{fig:flowdroid_cha_missed}
\end{figure}

We observe, in the ``Min'' column that for some apps, the minimum number of methods missed is low, e.g., 14 for FlowDroid-CHA. 
This could indicate that, for some apps, the static models appear to be sound, i.e., it does not miss many methods called dynamically.
To investigate we have plotted the proportion of methods missed per app successfully analyzed for FlowDroid-CHA in Figure~\ref{fig:flowdroid_cha_missed}.
We can see that indeed, for some apps, the proportion of methods missed is low (i.e., for 6 apps, less than 10\% of methods were missed).
However, the proportion of methods missed grows quickly and is critical for most apps.

\begin{figure}[ht]
    \centering
    \begin{adjustbox}{width=.9\columnwidth,center}
    \includegraphics{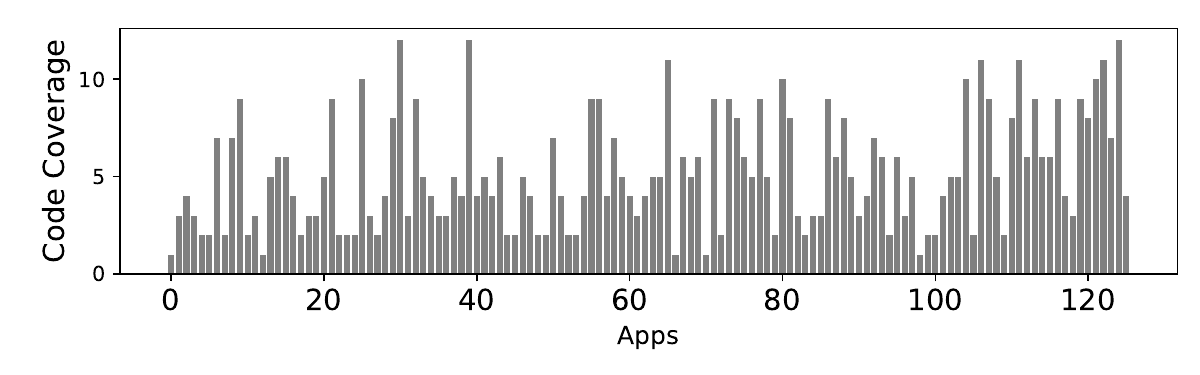}
    \end{adjustbox}
    \caption{Code Coverage of the dynamic analysis for the 126 apps successfully analyzed by all tools. The code coverage is at the method level and is expressed in \%.}
    \label{fig:126_code_coverage}
\end{figure}

Furthermore, Figure~\ref{fig:126_code_coverage} shows the proportion of code covered during the dynamic analysis for each of the 126 apps (in the same order as in Figure~\ref{fig:flowdroid_cha_missed}).
We see no correlation between unsoundness and code coverage, i.e., this is not because code coverage is better, that unsoundness is worse.

\begin{figure}[h!]
    \centering
    \begin{adjustbox}{width=.9\columnwidth,center}
        \includegraphics{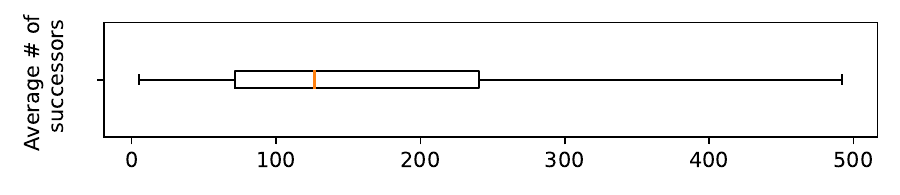}
    \end{adjustbox}
    \caption{Distributions of the average number of transitive successors of
        methods dynamically called.}
    \label{fig:avg_successor_both}
\end{figure}

Moreover, let us now focus on the dynamically-extracted call graph.
For each app, we have computed the following:
for each node in the call graph, the number of transitive successors
of the node.
Figure~\ref{fig:avg_successor_both} shows the distribution of the average number of transitive successors for each node for each app.
We can see that if a method $m$ is missed by a static analyzer, this is in fact hundreds of methods that will not be modeled, and not analyzed if $m$ is the only entrypoint to these methods.

\noindent
\textbf{Qualitative Analysis of the Methods Missed}
\label{sec:rq3:qualitative}
On our reduced dataset, there were 25 successful tool invocations on
  each of 126 apps = 3150 total invocations.
Table~\ref{tab:missed_methods} shows the top 10 most frequently missed
methods.
The top 5 methods belong to the com\-.ryan\-heise package, which is a Flutter plugin~\cite{audioservice}. 
Flutter is a framework for building apps.
The following 5 most frequently missed methods have their class name obfuscated, preventing us from identifying their origin.
In addition, 8 out of these 10 methods are class constructors. (In Java,
<init> refers to a constructor method, and <clinit> a static class initializer.)
The remaining 2 methods, onCreate()--a lifecycle method of the AudioService class that extends the MediaBrowserServiceCompat class--and size(), were called during run time but not modeled statically. 
This is unusual, given that static analyzers like FlowDroid are expected to handle lifecycle methods and other standard methods such as constructors.

\begin{table}[ht]
    \centering
    \caption{Top 10 most missed methods by static analyzers.}
    \label{tab:missed_methods}
    \begin{adjustbox}{width=.9\columnwidth,center}
    \begin{tabular}{|c|c|}
        \hline
        Occurrences & Method \\
        \hline
        2594 & com.ryanheise.audioservice.AudioService.<clinit>() \\
        2594 & com.ryanheise.audioservice.AudioService.<init>() \\
        2562 & com.ryanheise.audioservice.AudioService\$d.<init>(AudioService) \\
        2555 & com.ryanheise.audioservice.AudioService\$a.<init>(AudioService,int) \\
        2551 & com.ryanheise.audioservice.AudioService.onCreate() \\
        2422 & l.b.<init>() \\
        2358 & k.a.<init>() \\
        2315 & k.b.<init>() \\
        2290 & l.b.size() \\
        2290 & r.g.<init>() \\
        \hline
    \end{tabular}
    \end{adjustbox}
    
    \centering
    \caption{Top 10 most missed methods by static analyzers without obfuscated class names.}
    \label{tab:missed_methods_non_obfuscated}
    \begin{adjustbox}{width=.9\columnwidth,center}
    \begin{tabular}{|c|c|}
        \hline
        Occurrences & Method \\
        \hline
        2594 & com.ryanheise.audioservice.AudioService.<clinit>() \\
        2594 & com.ryanheise.audioservice.AudioService.<init>() \\
        2562 & com.ryanheise.audioservice.AudioService\$d.<init>(AudioService) \\
        2555 & com.ryanheise.audioservice.AudioService\$a.<init>(AudioService,int) \\
        2551 & com.ryanheise.audioservice.AudioService.onCreate() \\
        2203 & com.unity3d.player.h.onActivityStopped(Activity) \\
        2191 & com.unity3d.player.h.onActivityPaused(Activity) \\
        2125 & vn.hunghd.flutterdownloader.FlutterDownloaderInitializer.onCreate() \\
        2125 & vn.hunghd.flutterdownloader.FlutterDownloaderInitializer.a(Context) \\
        2098 & vn.hunghd.flutterdownloader.FlutterDownloaderInitializer\$a.<init>(g) \\
        \hline
    \end{tabular}
    \end{adjustbox}
\end{table}

Among the next 5 methods with
non-obfuscated class names (Table~\ref{tab:missed_methods_non_obfuscated}),
two methods come from the unity3d framework~\cite{unity}, and the other 3 belong to a Flutter plugin~\cite{flutterDownloader}. 
For all of the above 15 methods, some algorithms included them in the
  call graph.

\Cref{tab:missed_methods,tab:missed_methods_non_obfuscated} underscore
that frameworks present a significant obstacle to the precise static
modeling of Android
apps\cite{10.1145/2771284.2771288,10.1145/2884781.2884816}, with
quantitative data beyond previous investigations.

We have contacted the authors of all the tools considered in our study.
We have received 5 replies from the authors of RAICC, BackDroid, Gator, ArpDroid, and Difuzer.
None of them is surprised about the unsoundness of their model and say that the goal of their tool is not compute a sound and complete call graph.

\highlight{
\textbf{RQ2 answer:}
Our findings reveal an inverse relationship between the precision of a call graph construction algorithm and its soundness.
Although algorithms with less precision appear more sound, they too exhibit substantial soundness issues.
The proportion of methods missed by static analyzers in their call graph varies significantly, with CHA-based tools missing a minimal 40\% while SPARK algorithm-based tools exhibit substantial unsoundness, missing up to an astonishing \num{100}\% for SootFX.
}

\subsection{RQ3: Root Causes of Unsoundness}
\label{sec:cause_unsoundness}

\Cref{sec:rq2} showed that static modeling misses many methods --- that
  is, it is significantly unsound.  This section asks, why are they overlooked?

\textbf{Dynamic Call Graph and Entry Point Methods. }
Our dynamic analysis (\cref{sec:dynamic_analysis}) captures every call from
within the app, and every execution of every method in the app.
We remind we instrumented all apps to log every method call during execution, as detailed in Section~\ref{sec:dynamic_analysis}. 
Using the set of dynamically collected edges $DE$, this procedure generates a call graph for each app, call graph that we will call \emph{Dynamic Call Graph} (or DCG in short) to differentiate this call graph from the one obtained via static analysis.
If a
method is ever executed without being called from within the app, we call
it an \emph{entry point}.
\emph{We hypothesize that these entry point methods are a major cause of unsoundness.}

\begin{figure}[ht]
  \centering
    \includegraphics[width=.9\columnwidth]{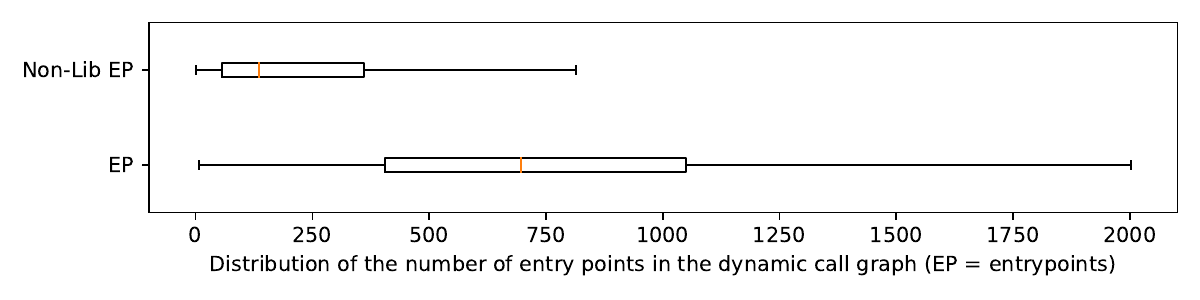}
  \caption{Distribution of the number of entry point methods called during the dynamic analysis.}
  \label{fig:distribution_entrypoint_methods}
\end{figure}

Figure~\ref{fig:distribution_entrypoint_methods} shows the distribution of
the number of entry point methods in the DCG for the \num{1000} apps of our
dataset (with and without considering libraries: note that libraries are embedded in Android apps when apps are packaged; thus libraries can be entry points).
When libraries are considered, the median is 697, representing \num{16.37}\% of the total number of methods in the DCG, and the mean is 833 per app.
For non-library methods, the median is 136, representing \num{4}\% of the total number of methods and 20.2\% of non-library methods.  
The mean is 284 per app. 
Both distributions are significantly different, as confirmed by a Mann-Whitney-Wilcoxon (MWW) test~\cite{10.1214/aoms/1177730491} (significance level set at 0.05).
Our results indicate that, on average, $1/6$ of the methods in the DCG are entry points, 
and a $1/5$ of non-library methods are entry points.
Also, this result indicates that most entry point methods are methods from libraries.

These entry point methods are key in our study because they are hard for static analyzers to discover. Indeed,  there is no call to these methods in the app code. So, without proper modeling of these entry points methods, a static analyzer will simply miss them. 
Moreover, as shown in Section~\ref{sec:rq2}, if one method is missed, there are hundreds of additional transitive methods potentially missed.

Let us now further investigate these entry point methods.
Table~\ref{tab:top_ten_entrypoint_dynamic} shows the top ten entry point methods (with libraries) in the DCGs of apps in our dataset.
Six methods are from the \texttt{androidx} package and four are from the \texttt{com.google} package, which provides additional libraries developed by Google.

\begin{table}[ht]
    \centering
    \caption{Top 10 entry point methods in the DCG of the 1000 apps (ranked per number of occurrences)}
    \label{tab:top_ten_entrypoint_dynamic}
    \begin{adjustbox}{width=.9\columnwidth,center}
    \begin{tabular}{|c|c|}
        \hline
        Occurrences & Method \\
        \hline
        961 & androidx.core.app.CoreComponentFactory.<init>() \\
        955 & androidx.core.app.CoreComponentFactory.instantiateApplication(ClassLoader,String) \\
        944 & androidx.core.app.CoreComponentFactory.instantiateProvider(ClassLoader,String) \\
        794 & androidx.startup.InitializationProvider.onCreate() \\
        790 & com.google.firebase.provider.FirebaseInitProvider.onCreate() \\
        786 & androidx.startup.InitializationProvider.<init>() \\
        781 & com.google.firebase.provider.FirebaseInitProvider.attachInfo(Context,ProviderInfo) \\
        780 & com.google.firebase.provider.FirebaseInitProvider.<init>() \\
        763 & androidx.core.app.CoreComponentFactory.instantiateActivity(ClassLoader,String,Intent) \\
        713 & com.google.android.gms.dynamite.DynamiteModule.<clinit>() \\
        \hline
    \end{tabular}
    \end{adjustbox}
\end{table}

\begin{table}[ht]
    \centering
    \caption{Top 10 entry point non-library methods in the 1000 dynamic call graphs}
    \label{tab:top_ten_entrypoint_dynamic_non_lib}
    \begin{adjustbox}{width=.9\columnwidth,center}
    \begin{tabular}{|c|c|}
        \hline
        Occurrences & Method \\
        \hline
        191 & com.unity3d.player.UnityPlayer\$5.run() \\
        185 & com.ryanheise.audioservice.AudioService.onCreate() \\
        185 & com.ryanheise.audioservice.AudioService.<init>() \\
        184 & com.ryanheise.audioservice.AudioService.<clinit>() \\
        184 & com.unity3d.player.UnityPlayer\$d.<clinit>() \\
        183 & com.unity3d.player.UnityPlayer\$1.onClick(DialogInterface,int) \\
        181 & k.a.<clinit>() \\
        174 & b1.a.<clinit>() \\
        170 & com.unity3d.player.UnityPlayer.<clinit>() \\
        168 & b1.b.<clinit>() \\
        \hline
    \end{tabular}
    \end{adjustbox}
\end{table}

\textbf{Entry Point Methods Missed by Static Analyzers:}
Previously, we have seen that many different methods from many classes are entry points in the DCG, but we have not yet checked if static analyzers miss these methods.
In fact, all the top 10 methods of Table~\ref{tab:top_ten_entrypoint_dynamic} and Table~\ref{tab:top_ten_entrypoint_dynamic_non_lib} are missed by static analyzers.
We have investigated further, and we observed that all static analyzers overlooked the top \num{5352} entry point methods among the \num{776075} entry points identified. 
Among these identified entry points, 
34.5\% (\num{267843}) were missed by static analyzers.
Regarding non-library entry point methods, representing \num{272313} of the total, the top \num{5863} methods were not modeled. 
In total, \num{95309} methods were missed, representing \num{35}\% of the non-library entry points.
Our results highlight a substantial gap in the coverage provided by static analysis tools concerning identifying entry points of the DCGs.

Note that these missed entry point methods represent \num{20.3}\%  of the total methods missed by the static analyzers.
The rest of the methods missed (i.e.,  \num{79.7}\%) are transitive methods of the dynamic entry points.
This result validates our hypothesis that entry points in the DCG are one of the main causes of unsoundness in Android app static analyzers.

\begin{figure}[ht]
  \centering
    \includegraphics[width=.9\columnwidth]{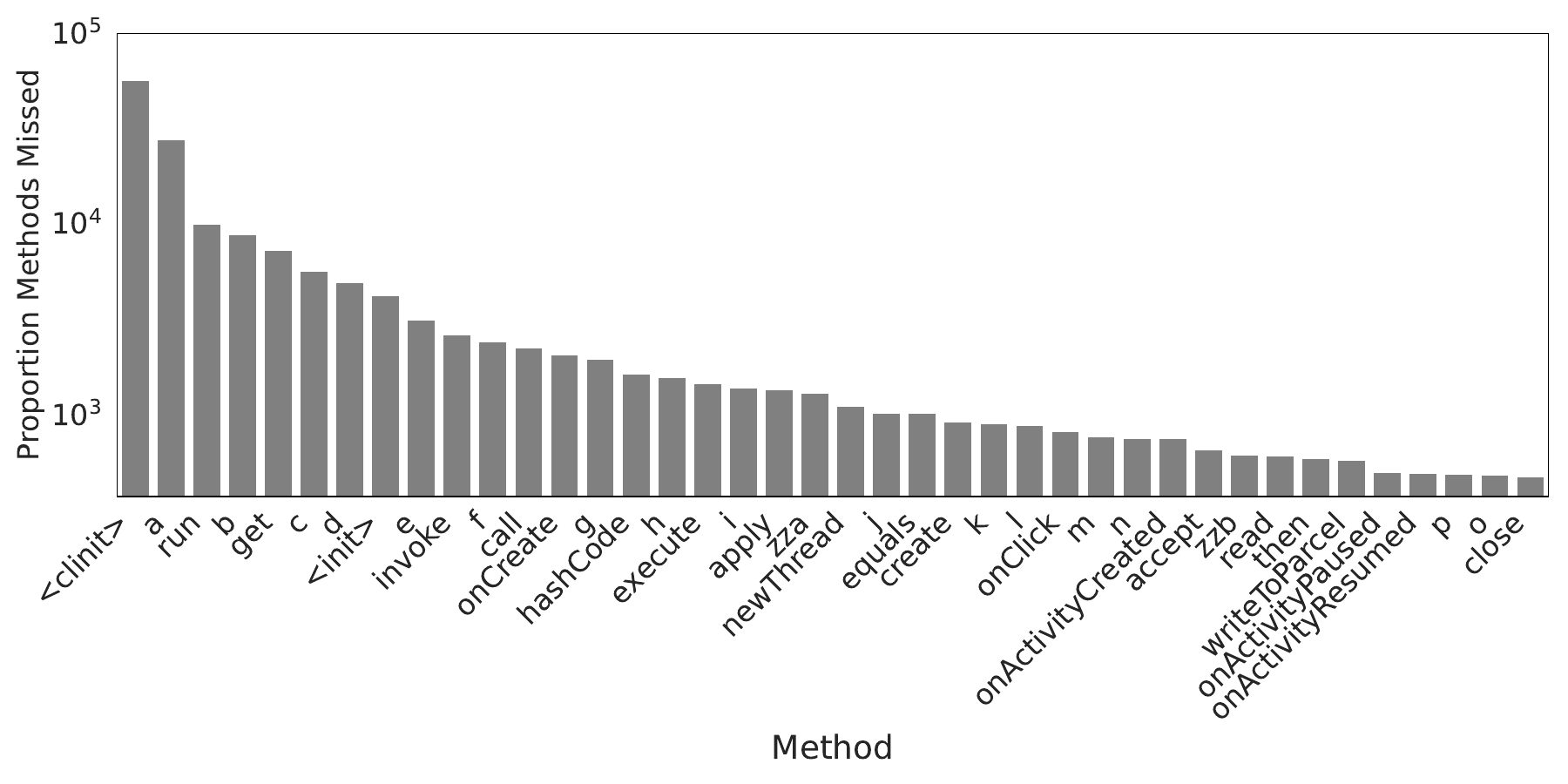}
  \caption{Top 40 most unique missed entry point method' names by static analyzers (logarithmic scale)}
  \label{fig:methods_missed_top_40}
\end{figure}

Let us now examine the top 40 method names that were overlooked by static analyzers among the entry points identified through dynamic analysis.
Note that, in this part, we only look at \emph{method names}, i.e., no matter their class (as studied in Tables~\ref{tab:top_ten_entrypoint_dynamic} and~\ref{tab:top_ten_entrypoint_dynamic_non_lib}).
Figure~\ref{fig:methods_missed_top_40} illustrates the number of occurrences of method names found in the set of dynamic entry points 
that were not captured by static analyzers.
First, we notice that static constructors (i.e., $<clinit>$) are by far the most missed methods with \num{56708} occurrences.
By comparison, constructors (i.e., $<init>$) have \num{4209} occurrences.
This result indicates that constructors are particularly prone to being overlooked by static analysis tools.
We observe that a high proportion of obfuscated methods are also hard to model by static analyzers (i.e., methods with names such as ``a'', or ``b'').
Interestingly, several methods that are indicative of well-known implicit mechanisms such as run(), call(), onCreate(), execute(), onClick(), etc., continue to be neglected by static analyzers.
This observation shows significant opportunities for improving the soundness of existing tools and techniques.

\begin{figure}[ht]
  \centering
    \includegraphics[width=.9\columnwidth]{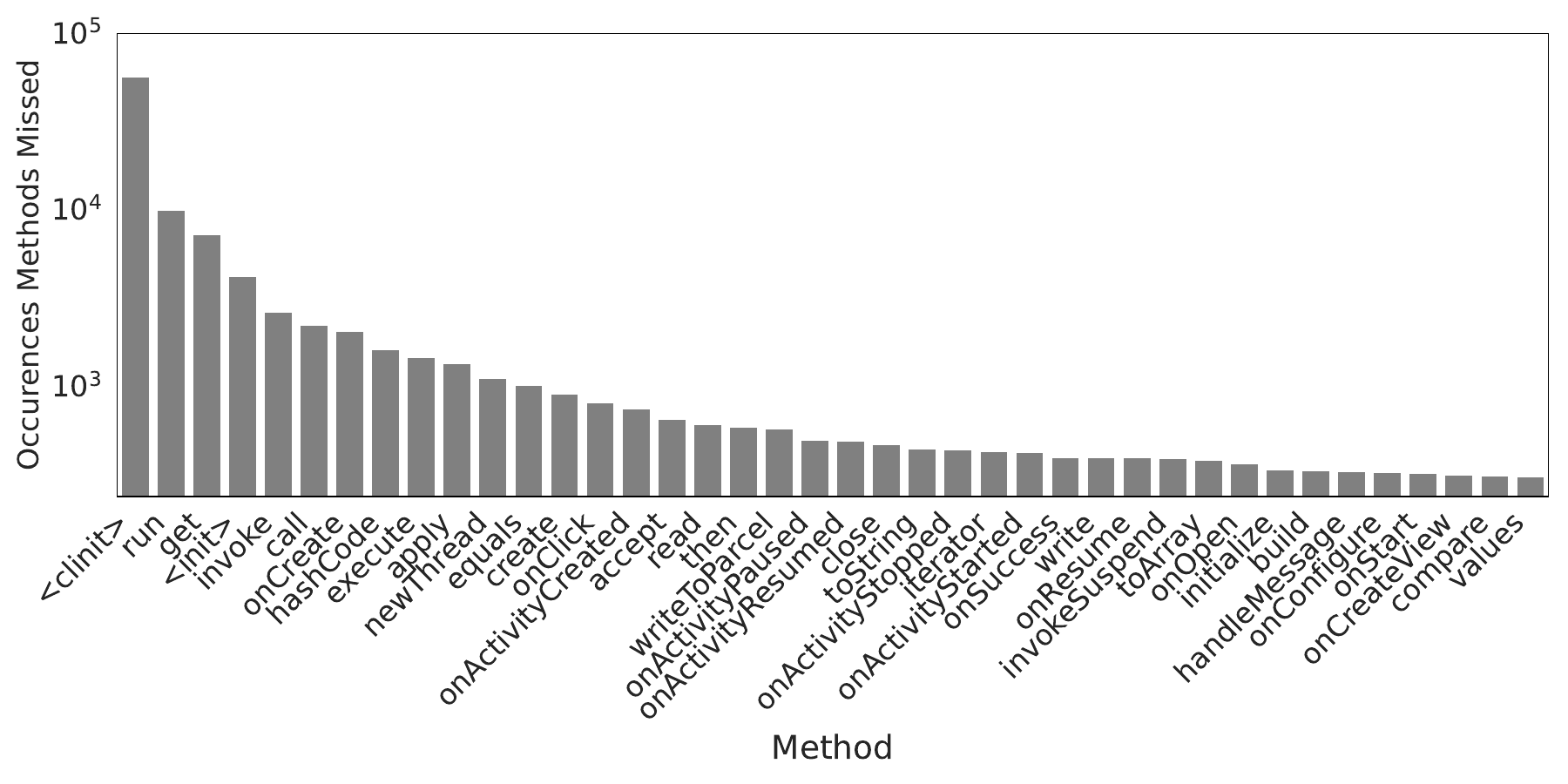}
  \caption{Top 40 most unique non-obfuscated missed entry point method' names by static analyzers}
  \label{fig:methods_missed_top_40_non_obfuscated}
\end{figure}

Further, Figure~\ref{fig:methods_missed_top_40_non_obfuscated} shows the same information as in Figure~\ref{fig:methods_missed_top_40} but without obfuscated method names.
We still observe well-known implicit mechanism with methods start with ``on'', such as on\-Con\-fig\-ure, on\-Re\-sume, etc.
But, we also see that several other methods are overlooked, such as accept, read, close, build, values, etc.
Our results give insights into which direction future research should dig into to improve the soundness of static analysis.
We provide, in our replication package, a list of \num{7137} unique method names and the list of \num{220127} methods (with their class) to provide insights for future research to improve the soundness of static analysis of Android apps.

\highlight{
\textbf{RQ3 answer:}

Determining the root causes of missed methods is complex, necessitating a thorough investigation of all missed methods in apps.
Despite this complexity, our study offers several crucial insights:
\dcircle{1} static analyzers lack understanding of the Android framework and external frameworks; and
\dcircle{2} modeling entry point methods challenge call graph construction.
Our study highlights many opportunities for future research and paves the way for improving the overall soundness of static analysis tools.
}

\subsection{Case Study}

This section presents examples of methods missed by static analysis tools.
These examples were selected randomly among the methods that were missed by static analyzers.

\textbf{Case 1:}
Method com.lyokone.location.Flutter\-Location\-Service.a\-(Ac\-tiv\-i\-ty)\footnote{\scriptsize in app: 78064E0B68067BC764102B47391F0D912F8C250E17A80FDC3828EBBEA53F497F} is invoked by h.j.a.b(), which is called by h.j.a.a(), and ultimately by h.j.a.on\-De\-tached\-From\-Activity(), a Flutter method called implicitly defined in class io.flutter.\-em\-bed\-ding\-.engine\-.plug\-ins\-.ac\-tiv\-i\-ty\-.Ac\-tiv\-i\-ty\-A\-ware.
To conclude, method com.lyokone.location.Flut\-ter\-Location\-Ser\-vice.a\-(Activity) is missed because of an \textbf{initial implicit call} that is not modeled since the h.j.a.on\-De\-tached\-From\-Activity() methods is an entry point in the DCG.

\textbf{Case 2:}
Method com.gulf\-brokers.an\-droid.app.database.My\-Data\-base\_Impl.create\-Invalidation\-Track\-er()\footnote{\scriptsize in app: E44DDFB0FDE572171BA60595B7AD6BC95AA7ACFA8AA932473C4AE6CBC0A3589C}  which is invoked in the static initializer of the abstract RoomDatabase class. 
MyDatabase\_\-Impl (a subclass of RoomDatabase) is instantiated in the onOpen() method, which is declared in an anonymous object created in the app that extends the androix.room.Room\-Open\-Help\-er.Del\-e\-gate class (which necessitates onOpen() to be overridden).
Method on\-Open() is called implicitly by androidx (part of the Android framework) and is, therefore, an entry point in the DCG.

\textbf{Case 3:}
The method c.a.a.a.i.x.j.h0.on\-Open(android.data\-base\-.sqlite\-.SQLite\-Data\-base)\footnote{\scriptsize in app: 14DDDDBCB6395363B490A32C33A8924E16F94295F77E2DC27A1453D754465ABC}, inherited from the android.database.sqlite\-.SQLiteOpenHelper class, which implements five implicit mechanisms, was not modeled successfully.
This method is called implictly
by the android framework.
Again, it is an entry point in the DCG.

Similarly, methods from the Cordova and Flutter frameworks were also missed.
Specifically, methods com.get\-ca\-pac\-i\-tor.cor\-dova.\-Mock\-Cor\-dovaWebViewImpl\$Ca\-pac\-i\-tor\-Eval\-Bridge\-Mode\-.on\-Na\-tive\-To\-Js\-Mes\-sage\-Available(org\-.apache\-.cor\-dova.Native\-To\-Js\-Mes\-s\-a\-g\-e\-Q\-u\-e\-u\-e)\footnote{\scriptsize in app: 6E2AB5488A78E61BF63EE4CFD942E85D92C2CB10F99F86E2338448E8346555D1} and co\-m.\-mr.\-flutter.plugin.file\-picker.FilePickerPlugin.on\-A\-t\-t\-a\-ched\-To\-En\-gi\-ne\-(io.flutter.em\-bed\-ding.\-engine.plugins.Flut\-ter\-Plug\-in\-\$\-F\-l\-ut\-ter\-Plugin\-Bind\-ing)\footnote{\scriptsize in app: 78064E0B68067BC764102B47391F0D912F8C250E17A80FDC3828EBBEA53F497F}. 
These methods, called implicitly by
  the respective frameworks, are \textbf{never called
within the apps}, which prevented the tools from modeling them.

These examples indicate clear limitations into handling implicit mechanisms across various frameworks from existing static analysis tools.
This suggests a research gap, indicating that further investigation into these frameworks is needed. 
By studying these frameworks and identifying their implicit mechanisms, there is a potential to improve static modeling significantly. 
Moreover, our research suggests that the Android framework still contain numerous unexplored implicit mechanisms, highlighting the necessity for continued exploration and analysis.

\subsection{RQ4: Can Systematically-Collected Callbacks Improve Soundness?} 

To the best of our knowledge, two papers, in which the authors devise techniques to collect implicit mechanisms from the Android Framework, have been presented.
The first one presents EdgeMiner~\cite{cao2015edgeminer}, a technique to collect callbacks (i.e., a type of implicitly invoked method) systematically from the Android framework.
Their technique involves the static analysis of the Android framework in order to generate summaries of API methods describing the implicit control flow transitions of callbacks.
The second one presents Columbus~\cite{columbus} an automated technique that statically analyzes the Android framework and apps under test.
Their technique involves identifying apps' methods that override framework methods to build a mapping between registration and callback methods.

To determine whether systematic approaches aiming to capture implicit mechanisms in the Android framework cover the methods missed by the static analysis tools studied, we have contacted the authors of Edge\-Miner and Columbus.
Unfortunately, we could not gather data from Columbus due to a lack of response and an empty repository~\cite{columbusrepository}. 
Edge\-Miner's authors shared their data, allowing us to conduct a comparative analysis.

By extracting potential callback methods from the Edge\-Miner dataset and comparing these with the methods called dynamically in our study, we could evaluate the potential improvement Edge\-Miner would bring to static analyzers.
We have computed the difference between the set of unique called methods ($\|DM|$) and the set of Edge\-Miner's unique potential callbacks ($EM$), expressed as $\|DM| \setminus EM$.

With $|\|DM||$ being \num{1082265} and $|EM|$ being \num{19510}, we found that $|\|DM| \setminus EM|$ equaled \num{1081886}. 
This suggests that Edge\-Miner would only help static analyzers to model
  \num{1082265}\,$-$\,\num{1081886} =
379 additional methods out of the \num{1082265} called dynamically. 
Our research suggests that the findings reported in the EdgeMiner's paper may not hold the same level of efficiency for improving the soundness of static analyzers.
This limited improvement that Edge\-Miner would bring to static analyzers might be attributed to several factors: 
\dcircle{1} The potential presence of false-positive results in the Edge\-Miner dataset (i.e., methods collected not being callbacks); and
\dcircle{2} The fact that Edge\-Miner focuses on the Android Framework, thereby potentially missing callbacks from external frameworks like React Native, or Flutter, as well as those triggered by languages such as C++ or JavaScript, which can also trigger bytecode methods.

\highlight{
\textbf{RQ4 answer:}
Existing approaches that systematically collect implicit mechanisms from the Android framework dot not significantly augment the static models of current static analyzers.
}
\section{Threats to validity}
\label{sec:limitations}

As discussed in \cref{sec:dynamic-analysis-low-coverage},
the runs in our experiments achieve low coverage.
This means that our results underestimate the true extent of the missed methods. 

Our evaluation is over only \numtools static analysis tools and 25
  configurations.
  This was all the open-source tools we could find and run after extensive
  efforts.  Other tools may behave differently, but \numtools tools already offer
substantial breadth to our investigation.

Our study's scope was limited to Dalvik bytecode, which is the
compiled form of Java/Kotlin Android code and is the dominant form of
  code that runs in Android apps.
If an app used other languages e.g., C or JavaScript, they would be treated
    by our experiment like library code. 
To the best of our knowledge, among the tools considered in our study, only NatiDroid models C code to get a more accurate specification of Android API protection.
When a static tool does not model C and JavaScript code, its unsoundness is
greater than reported in our experiments.

Static analysis might be hindered by packed apps.
Packing is a technique to obscure code or data. 
The process involves compressing or encrypting code, which is uncompressed or decrypted at run time. 
This makes static analysis more difficult since the code is hidden and only revealed during execution.
To check for packing, we used the ApkId tool~\cite{apkid}.
It was observed that only two apps out of the 1000 apps of our dataset
were using packed code.
In both cases, the packer used was
DexProtector~\cite{dexprotextor}, it is used to protect apps from tampering, reverse-engineering, and cracking.
As a result,
since only 2 apps use packing, we can eliminate packing as a potential barrier to creating sound call graphs in our dataset.

Similarly, static analysis can be hindered by dynamic loading.
For instance, classes might be downloaded from an external server and loaded dynamically at run time.
In this case, existing static analyzers cannot account for dynamically loaded code.
Note that packing and dynamic loading can also affect our dynamic analysis since we cannot instrument the packed or loaded code dynamically.
This is mitigated by the fact that, if we had instrumented more methods, we would have shown that the problem of unsoundness is worse than it appears.

The performance of a static analysis is affected by its configuration.
We used each tool's default configuration, but it is possible that some
other configuration would have yielded better call graphs.

Finally, the 1-hour timeout to compute a call graph may not be adequate for some apps or tools, as increasing the timeout could allow to fully analyze more apps.
\section{Related Work}
\label{sec:related_work}

The literature contains many approaches to handle implicit mechanisms for the Android platform.
\Cref{sec:related-implicit} describes studies that only handle particular
implicit mechanisms in Android apps and propose static model to account for
them.
\Cref{sec:related-systematic} presents studies that aim at systematically analyze the Android framework to find implicit mechanisms.
\Cref{sec:related-soundness} presents studies that have attempted to measure call graph soundness.

\subsection{Handling particular implicit mechanisms}
\label{sec:related-implicit}

\noindent
\textbf{Callbacks.}
Callback mechanisms refer to the registration of a method executed by the Android framework in response to events, e.g., a click. 
The methods triggered are then called implicitly by the framework, i.e., there is no explicit call to them in the app code which poses a challenge to static analyzers.
Therefore, techniques have been devised in the literature to account for callbacks.

FlowDroid~\cite{10.1145/2666356.2594299} was a pioneer in statically modeling callbacks in Android apps.
It would construct a call graph per component, identifying calls to system methods with callback interfaces (defined in layout XML files) and incrementally extending the call graph with newly added method calls.
Furthermore, FlowDroid also includes handpicked callback methods in configuration files.
Yang et al.~\cite{10.1109/ICSE.2015.31} conducted a study on lifecycle and user-driven callbacks in Android apps. 
Their approach utilizes a GUI model to capture the app's graphical user interface and generates a callback control flow graph. 
By analyzing the generated GUI model, the authors extract possible sequences of user GUI events that correspond to valid paths in the model.
Likewise, Wu et al.~\cite{10.1109/TSE.2016.2547385} introduced a callback-aware technique focusing on two callback methods: system-triggered and user-triggered. 
The former includes lifecycle and callback methods of resource classes, such as onDestroy(), while the latter represents callbacks triggered by user interactions with the GUI.

\noindent
\textbf{Inter-Component Communication.}
Android apps consist of various components that communicate with each other through inter-component communication (ICC) methods provided by the Android framework, such as startActivity() and sendBroadcast(). 
These ICC methods trigger the execution of lifecycle methods implemented by each component, including onCreate() and on\-Re\-ceive(). 
The communication between components through ICC methods involves implicit calls to these lifecycle methods from the Android framework.
Numerous research efforts have been dedicated to resolving the target components of ICC communication. 

IccTA~\cite{10.5555/2818754.2818791} relies on composite constant propagation~\cite{10.5555/2818754.2818767} and instrumentation to infers the potential targets of \texttt{Intents}, while Amandroid~\cite{10.1145/2660267.2660357} generates data flow and data dependence graphs to infer possible target components.
DroidSafe~\cite{gordon2015information} employs string and class analysis to infer target components and modifies ICC method calls to explicit lifecycle method calls.
RAICC~\cite{10.1109/ICSE43902.2021.00126} addresses atypical ICC methods, such as SmsManager.sendTextMessage(), by resolving potential targets with constant propagation and instrumenting the app to include ICC method calls (e.g., startActivity()).
ICCBot~\cite{10.1109/ICSE-Companion55297.2022.9793791} is a recently released tool that performs context-sensitive and inter-procedural analysis to infer component transitions connected via fragments, modeling data carried by ICC objects like Intents.
Additionally, Chen et al.~\cite{chen2022automatically,10.1109/ICSE.2019.00070} also recently developed an approach to construct an Activity Transition Graph to create storyboards for apps using ICC-related information to improve activity coverage.

\noindent
\textbf{Reflection.}
The reflection mechanism allows for run-time introspection, enabling the execution of methods without explicitly calling them in the source code, i.e., using reflective calls.

More than ten years ago, TamiFlex~\cite{10.1145/1985793.1985827} was proposed to boost static analyzers with information dynamically gathered (thanks to instrumentation) about reflective calls.
TamiFlex also introduced an instrumentation engine to insert regular method calls into apps to boost existing static analyzers.
DroidRA~\cite{10.1145/2931037.2931044} is an instrumentation-based analysis technique that enhances Android apps by resolving reflective calls using the COAL solver to infer reflection targets. 
By instrumenting the app and adding explicit calls for each resolved reflective call, DroidRA improves analysis accuracy. 
In addition, Barros et al.~\cite{7372054} propose a two-fold solution for resolving Java reflection call targets. 
Their approach includes a reflection-type system for inferring class and method names and a reflection solver for estimating invocable method signatures.

\ding{224} Contrary to these approaches, we do not aim at identifying and focusing on single implicit mechanisms in Android apps.
Our empirical study has been devised to capture any method call --implicit or not-- happening at execution time but not modeled statically to understand and thus improve static analysis models.

\subsection{Systematic studies}
\label{sec:related-systematic}

\noindent
\textbf{EdgeMiner.}
Edge\-Miner~\cite{cao2015edgeminer} systematically analyzes multiple versions of the Android framework to identify callbacks and their corresponding registration methods.
Edge\-Miner relies on inter-procedural backward data flow analysis to detect and link callbacks to their registration methods.

\noindent
\textbf{Columbus.}
The latest work that systematically analyzes the Android framework to search for callbacks is Columbus~\cite{columbus}.
The authors statically analyze the Android framework to identify callbacks by considering protected or public methods with at least one caller in the framework. 
Then they construct a call graph of the framework and over-approximates possible targets when type inference fails. 
Additionally, the authors consolidate callbacks inherited from superclasses by traversing the class hierarchy to produce a mapping from the registration method to the callback method.

\ding{224} Contrary to these works, our study aims at revealing the extent to which static analyzers' model are unsound with a high level a precision.
Also, our work do not focus on the Android framework and \emph{callbacks},
rather it encompasses any method that could be missed statically, and from
any source, e.g., an external framework.

\subsection{Call graph soudness}
\label{sec:related-soundness}

Ali et al.~\cite{8944149} investigate the efficacy of JVM-bytecode-based static analysis across various JVM-hosted languages, including Scheme, Scala, OCaml, Groovy, Clojure, Python, and Ruby.
The authors found that the analyses produce sound call graphs for Scheme, Scala, and OCaml, similar to Java, but fails to do so for Groovy, Clojure, Python, and Ruby due to their extensive use of reflection and invokedynamic instructions.
Reif et al.~\cite{10.1145/3293882.3330555} have presented Judge, a toolchain designed to identify sources of unsoundness in call graphs.
The authors leverage Judge and a test suite to compare different call graph implementations (Soot, WALA, DOOP, and OPAL), evaluate language features and APIs' prevalence that impact soundness in modern Java bytecode.
Sui et al.~\cite{10.1007/978-3-030-02768-1_4} study the prevalence of dynamic language features in modern programming languages. 
It catalogs dynamic features for Java and presents a micro-benchmark that helps investigate the soundness of static analysis framework (i.e., Soot, Wala, and Doop).
Aljawder~\cite{https://open.bu.edu/handle/2144/17085} compares static call
graphs built with FlowDroid (using an unspecified call graph construction
algorithm) with dynamic call graphs resulting from execution with Monkey.
They found that 62/92 apps were missing edges in the static
call graph, but did not investigate root causes.
Wang et al.~\cite{10.1145/2931021.2931026} address the unsoundness in static analysis of Android GUIs, highlighting mismatches between existing tools like FlowDroid, IccTA, GATOR, and runtime behavior. 
Their study shows that these tools often miss runtime sequences of callbacks and parameters, due to the complexity of the Android framework. 
The paper suggests enhancing static analysis by incorporating insights from runtime behavior to reduce unsoundness.

\ding{224} Contrary to these studies, we do not concentrate on specific language features.
We examine real-world call graphs produced by existing static analysis tools and dynamic analysis.
We then compare these call graphs, highlighting the discrepancies among them, i.e., methods called at runtime that \emph{must} appear in the static call graph but that do not.
Our study reveals that many methods are missed in real-world apps, not only due to language features, such as those studied in these papers, but due to, e.g., implicit calls.
\section{Conclusion}
\label{sec:conclusion}

We have devised an empirical study to measure the degree to which static analyzers under-approximate their static models and aimed to uncover the sources of the discrepancies between static models and dynamically collected data. 
Our investigation offered several key insights: firstly, both Android and external frameworks pose significant challenges to the current state-of-the-art static analyzers when attempting to build sound models. 
Secondly, call graph algorithms with high precision levels result in a high degree of unsoundness. 
Thirdly, no existing systematic or specific approach drastically ameliorates soundness. 
Finally, we identified that modeling entry point methods significantly challenge call graph construction and soundness.
Our findings show that 
\dcircle{1} having the most effective and precise static analysis run over unsound static models of apps is useless; and
\dcircle{2} the problem we highlight in the paper is likely worse than described since the proportion of code covered during our dynamic analysis is low.
Therefore, there is a need for innovative techniques to improve the soundness of Android static analysis.
\section{Data Availability}
\label{sec:data_availability}

To promote transparency and facilitate reproducibility, our artifacts are
publicly available:
\begin{center}
    \url{https://github.com/JordanSamhi/Call-Graph-Soundness-in-Android-Static-Analysis}
\end{center}

\newpage

\bibliographystyle{ACM-Reference-Format}
\bibliography{bib}


\begin{thebibliography}{59}


\ifx \showCODEN    \undefined \def \showCODEN     #1{\unskip}     \fi
\ifx \showDOI      \undefined \def \showDOI       #1{#1}\fi
\ifx \showISBNx    \undefined \def \showISBNx     #1{\unskip}     \fi
\ifx \showISBNxiii \undefined \def \showISBNxiii  #1{\unskip}     \fi
\ifx \showISSN     \undefined \def \showISSN      #1{\unskip}     \fi
\ifx \showLCCN     \undefined \def \showLCCN      #1{\unskip}     \fi
\ifx \shownote     \undefined \def \shownote      #1{#1}          \fi
\ifx \showarticletitle \undefined \def \showarticletitle #1{#1}   \fi
\ifx \showURL      \undefined \def \showURL       {\relax}        \fi
\providecommand\bibfield[2]{#2}
\providecommand\bibinfo[2]{#2}
\providecommand\natexlab[1]{#1}
\providecommand\showeprint[2][]{arXiv:#2}

\bibitem[Ali et~al\mbox{.}(2021)]%
        {8944149}
\bibfield{author}{\bibinfo{person}{Karim Ali}, \bibinfo{person}{Xiaoni Lai},
  \bibinfo{person}{Zhaoyi Luo}, \bibinfo{person}{Ondřej Lhoták},
  \bibinfo{person}{Julian Dolby}, {and} \bibinfo{person}{Frank Tip}.}
  \bibinfo{year}{2021}\natexlab{}.
\newblock \showarticletitle{A Study of Call Graph Construction for JVM-Hosted
  Languages}.
\newblock \bibinfo{journal}{\emph{IEEE Transactions on Software Engineering}}
  \bibinfo{volume}{47}, \bibinfo{number}{12} (\bibinfo{year}{2021}),
  \bibinfo{pages}{2644--2666}.
\newblock
\urldef\tempurl%
\url{https://doi.org/10.1109/TSE.2019.2956925}
\showDOI{\tempurl}


\bibitem[Aljawder(2016)]%
        {https://open.bu.edu/handle/2144/17085}
\bibfield{author}{\bibinfo{person}{Dana Aljawder}.}
  \bibinfo{year}{2016}\natexlab{}.
\newblock \emph{\bibinfo{title}{Identifying unsoundness of call graphs in
  {Android} static analysis tools}}.
\newblock \bibinfo{thesistype}{Master's\ thesis}. \bibinfo{school}{Boston
  University}.
\newblock
\urldef\tempurl%
\url{https://open.bu.edu/handle/2144/17085}
\showURL{%
\tempurl}


\bibitem[Allix et~al\mbox{.}(2016)]%
        {10.1145/2901739.2903508}
\bibfield{author}{\bibinfo{person}{Kevin Allix},
  \bibinfo{person}{Tegawend{\'e}~F. Bissyand{\'e}}, \bibinfo{person}{Jacques
  Klein}, {and} \bibinfo{person}{Yves Le~Traon}.}
  \bibinfo{year}{2016}\natexlab{}.
\newblock \showarticletitle{AndroZoo: Collecting Millions of Android Apps for
  the Research Community}. In \bibinfo{booktitle}{\emph{Proceedings of the 13th
  International Conference on Mining Software Repositories}} (Austin, Texas)
  \emph{(\bibinfo{series}{MSR '16})}. \bibinfo{publisher}{ACM},
  \bibinfo{address}{New York, NY, USA}, \bibinfo{pages}{468--471}.
\newblock
\showISBNx{978-1-4503-4186-8}
\urldef\tempurl%
\url{https://doi.org/10.1145/2901739.2903508}
\showDOI{\tempurl}


\bibitem[ApkID(2023)]%
        {apkid}
\bibfield{author}{\bibinfo{person}{ApkID}.} \bibinfo{year}{2023}\natexlab{}.
\newblock \bibinfo{title}{\url{https://github.com/rednaga/APKiD}}.
\newblock
\newblock
\newblock
\shownote{Accessed December 2023}.


\bibitem[Arzt and Bodden(2016)]%
        {10.1145/2884781.2884816}
\bibfield{author}{\bibinfo{person}{Steven Arzt} {and} \bibinfo{person}{Eric
  Bodden}.} \bibinfo{year}{2016}\natexlab{}.
\newblock \showarticletitle{StubDroid: Automatic Inference of Precise Data-Flow
  Summaries for the Android Framework}. In
  \bibinfo{booktitle}{\emph{Proceedings of the 38th International Conference on
  Software Engineering}} (Austin, Texas) \emph{(\bibinfo{series}{ICSE '16})}.
  \bibinfo{publisher}{Association for Computing Machinery},
  \bibinfo{address}{New York, NY, USA}, \bibinfo{pages}{725–735}.
\newblock
\showISBNx{9781450339001}
\urldef\tempurl%
\url{https://doi.org/10.1145/2884781.2884816}
\showDOI{\tempurl}


\bibitem[Arzt et~al\mbox{.}(2014)]%
        {10.1145/2666356.2594299}
\bibfield{author}{\bibinfo{person}{Steven Arzt}, \bibinfo{person}{Siegfried
  Rasthofer}, \bibinfo{person}{Christian Fritz}, \bibinfo{person}{Eric Bodden},
  \bibinfo{person}{Alexandre Bartel}, \bibinfo{person}{Jacques Klein},
  \bibinfo{person}{Yves Le~Traon}, \bibinfo{person}{Damien Octeau}, {and}
  \bibinfo{person}{Patrick McDaniel}.} \bibinfo{year}{2014}\natexlab{}.
\newblock \showarticletitle{FlowDroid: Precise Context, Flow, Field,
  Object-Sensitive and Lifecycle-Aware Taint Analysis for Android Apps}.
\newblock \bibinfo{journal}{\emph{ACM SIGPLAN NOTICES}} \bibinfo{volume}{49},
  \bibinfo{number}{6} (\bibinfo{date}{June} \bibinfo{year}{2014}),
  \bibinfo{pages}{259–269}.
\newblock
\showISSN{0362-1340}
\urldef\tempurl%
\url{https://doi.org/10.1145/2666356.2594299}
\showDOI{\tempurl}


\bibitem[Barros et~al\mbox{.}(2015a)]%
        {7372054}
\bibfield{author}{\bibinfo{person}{Paulo Barros}, \bibinfo{person}{Ren\'{e}
  Just}, \bibinfo{person}{Suzanne Millstein}, \bibinfo{person}{Paul Vines},
  \bibinfo{person}{Werner Dietl}, \bibinfo{person}{Marcelo d'Amorim}, {and}
  \bibinfo{person}{Michael~D. Ernst}.} \bibinfo{year}{2015}\natexlab{a}.
\newblock \showarticletitle{Static Analysis of Implicit Control Flow: Resolving
  Java Reflection and Android Intents}. In
  \bibinfo{booktitle}{\emph{Proceedings of the 30th IEEE/ACM International
  Conference on Automated Software Engineering}} (Lincoln, Nebraska)
  \emph{(\bibinfo{series}{ASE '15})}. \bibinfo{publisher}{IEEE Press},
  \bibinfo{pages}{669–679}.
\newblock
\showISBNx{9781509000241}
\urldef\tempurl%
\url{https://doi.org/10.1109/ASE.2015.69}
\showDOI{\tempurl}


\bibitem[Barros et~al\mbox{.}(2015b)]%
        {BarrosJMVDdAE2015}
\bibfield{author}{\bibinfo{person}{Paulo Barros}, \bibinfo{person}{Ren{\'e}
  Just}, \bibinfo{person}{Suzanne Millstein}, \bibinfo{person}{Paul Vines},
  \bibinfo{person}{Werner Dietl}, \bibinfo{person}{Marcelo d'Amorim}, {and}
  \bibinfo{person}{Michael~D. Ernst}.} \bibinfo{year}{2015}\natexlab{b}.
\newblock \showarticletitle{Static analysis of implicit control flow: Resolving
  {Java} reflection and {Android} intents}. In
  \bibinfo{booktitle}{\emph{Proceedings of the International Conference on
  Automated Software Engineering (ASE)}}. \bibinfo{address}{Lincoln, NE, USA},
  \bibinfo{pages}{669--679}.
\newblock


\bibitem[Blackshear et~al\mbox{.}(2015)]%
        {10.1145/2771284.2771288}
\bibfield{author}{\bibinfo{person}{Sam Blackshear}, \bibinfo{person}{Alexandra
  Gendreau}, {and} \bibinfo{person}{Bor-Yuh~Evan Chang}.}
  \bibinfo{year}{2015}\natexlab{}.
\newblock \showarticletitle{Droidel: A General Approach to Android Framework
  Modeling}. In \bibinfo{booktitle}{\emph{Proceedings of the 4th ACM SIGPLAN
  International Workshop on State Of the Art in Program Analysis}} (Portland,
  OR, USA) \emph{(\bibinfo{series}{SOAP 2015})}.
  \bibinfo{publisher}{Association for Computing Machinery},
  \bibinfo{address}{New York, NY, USA}, \bibinfo{pages}{19–25}.
\newblock
\showISBNx{9781450335850}
\urldef\tempurl%
\url{https://doi.org/10.1145/2771284.2771288}
\showDOI{\tempurl}


\bibitem[Bodden et~al\mbox{.}(2011)]%
        {10.1145/1985793.1985827}
\bibfield{author}{\bibinfo{person}{Eric Bodden}, \bibinfo{person}{Andreas
  Sewe}, \bibinfo{person}{Jan Sinschek}, \bibinfo{person}{Hela Oueslati}, {and}
  \bibinfo{person}{Mira Mezini}.} \bibinfo{year}{2011}\natexlab{}.
\newblock \showarticletitle{Taming Reflection: Aiding Static Analysis in the
  Presence of Reflection and Custom Class Loaders}. In
  \bibinfo{booktitle}{\emph{Proceedings of the 33rd International Conference on
  Software Engineering}} (Waikiki, Honolulu, HI, USA)
  \emph{(\bibinfo{series}{ICSE '11})}. \bibinfo{publisher}{Association for
  Computing Machinery}, \bibinfo{address}{New York, NY, USA},
  \bibinfo{pages}{241–250}.
\newblock
\showISBNx{9781450304450}
\urldef\tempurl%
\url{https://doi.org/10.1145/1985793.1985827}
\showDOI{\tempurl}


\bibitem[Bose et~al\mbox{.}(2023)]%
        {columbus}
\bibfield{author}{\bibinfo{person}{Priyanka Bose}, \bibinfo{person}{Dipanjan
  Das}, \bibinfo{person}{Saastha Vasan}, \bibinfo{person}{Sebastiano Mariani},
  \bibinfo{person}{Ilya Grishchenko}, \bibinfo{person}{Andrea Continella},
  \bibinfo{person}{Antonio Bianchi}, \bibinfo{person}{Christopher Kruegel},
  {and} \bibinfo{person}{Giovanni Vigna}.} \bibinfo{year}{2023}\natexlab{}.
\newblock \showarticletitle{COLUMBUS: Android App Testing Through Systematic
  Callback Exploration}. In \bibinfo{booktitle}{\emph{Proceedings of the
  International Conference on Software Engineering (ICSE)}}.
\newblock
\newblock
\shownote{45th International Conference on Software Engineering, ICSE 2023}.


\bibitem[Cao et~al\mbox{.}(2015)]%
        {cao2015edgeminer}
\bibfield{author}{\bibinfo{person}{Yinzhi Cao}, \bibinfo{person}{Yanick
  Fratantonio}, \bibinfo{person}{Antonio Bianchi}, \bibinfo{person}{Manuel
  Egele}, \bibinfo{person}{Christopher Kruegel}, \bibinfo{person}{Giovanni
  Vigna}, {and} \bibinfo{person}{Yan Chen}.} \bibinfo{year}{2015}\natexlab{}.
\newblock \showarticletitle{EdgeMiner: Automatically Detecting Implicit Control
  Flow Transitions through the Android Framework.}. In
  \bibinfo{booktitle}{\emph{NDSS}}.
\newblock


\bibitem[Chen et~al\mbox{.}(2022)]%
        {chen2022automatically}
\bibfield{author}{\bibinfo{person}{Sen Chen}, \bibinfo{person}{Lingling Fan},
  \bibinfo{person}{Chunyang Chen}, {and} \bibinfo{person}{Yang Liu}.}
  \bibinfo{year}{2022}\natexlab{}.
\newblock \showarticletitle{Automatically Distilling Storyboard with Rich
  Features for Android Apps}.
\newblock \bibinfo{journal}{\emph{IEEE Transactions on Software Engineering}}
  (\bibinfo{year}{2022}).
\newblock


\bibitem[Chen et~al\mbox{.}(2019)]%
        {10.1109/ICSE.2019.00070}
\bibfield{author}{\bibinfo{person}{Sen Chen}, \bibinfo{person}{Lingling Fan},
  \bibinfo{person}{Chunyang Chen}, \bibinfo{person}{Ting Su},
  \bibinfo{person}{Wenhe Li}, \bibinfo{person}{Yang Liu}, {and}
  \bibinfo{person}{Lihua Xu}.} \bibinfo{year}{2019}\natexlab{}.
\newblock \showarticletitle{StoryDroid: Automated Generation of Storyboard for
  Android Apps}. In \bibinfo{booktitle}{\emph{2019 IEEE/ACM 41st International
  Conference on Software Engineering (ICSE)}}. \bibinfo{pages}{596--607}.
\newblock
\urldef\tempurl%
\url{https://doi.org/10.1109/ICSE.2019.00070}
\showDOI{\tempurl}


\bibitem[DexProtector(2023)]%
        {dexprotextor}
\bibfield{author}{\bibinfo{person}{DexProtector}.}
  \bibinfo{year}{2023}\natexlab{}.
\newblock \bibinfo{title}{\url{https://dexprotector.com}}.
\newblock
\newblock
\newblock
\shownote{Accessed December 2023}.


\bibitem[Dilhara et~al\mbox{.}(2018)]%
        {10.1145/3197231.3197255}
\bibfield{author}{\bibinfo{person}{Malinda Dilhara}, \bibinfo{person}{Haipeng
  Cai}, {and} \bibinfo{person}{John Jenkins}.} \bibinfo{year}{2018}\natexlab{}.
\newblock \showarticletitle{Automated Detection and Repair of Incompatible Uses
  of Runtime Permissions in Android Apps}. In
  \bibinfo{booktitle}{\emph{Proceedings of the 5th International Conference on
  Mobile Software Engineering and Systems}} (Gothenburg, Sweden)
  \emph{(\bibinfo{series}{MOBILESoft '18})}. \bibinfo{publisher}{Association
  for Computing Machinery}, \bibinfo{address}{New York, NY, USA},
  \bibinfo{pages}{67–71}.
\newblock
\showISBNx{9781450357128}
\urldef\tempurl%
\url{https://doi.org/10.1145/3197231.3197255}
\showDOI{\tempurl}


\bibitem[Downloader(2023)]%
        {flutterDownloader}
\bibfield{author}{\bibinfo{person}{Flutter Downloader}.}
  \bibinfo{year}{2023}\natexlab{}.
\newblock
  \bibinfo{title}{\url{https://github.com/fluttercommunity/flutter_downloader}}.
\newblock
\newblock
\newblock
\shownote{Accessed December 2023}.


\bibitem[et~al.(2023)]%
        {reunify}
\bibfield{author}{\bibinfo{person}{Yonghui~Liu et al.}}
  \bibinfo{year}{2023}\natexlab{}.
\newblock \showarticletitle{ReuNify: A Step Towards Whole Program Analysis for
  React Native Android Apps}. In \bibinfo{booktitle}{\emph{Proceedings of the
  38th ACM/IEEE International Conference on Automated Software Engineering}}
  (Luxembourg, France) \emph{(\bibinfo{series}{ASE 2023})}.
  \bibinfo{publisher}{Association for Computing Machinery}.
\newblock


\bibitem[Feng et~al\mbox{.}(2014)]%
        {10.1145/2635868.2635869}
\bibfield{author}{\bibinfo{person}{Yu Feng}, \bibinfo{person}{Saswat Anand},
  \bibinfo{person}{Isil Dillig}, {and} \bibinfo{person}{Alex Aiken}.}
  \bibinfo{year}{2014}\natexlab{}.
\newblock \showarticletitle{Apposcopy: Semantics-Based Detection of Android
  Malware through Static Analysis}. In \bibinfo{booktitle}{\emph{Proceedings of
  the 22nd ACM SIGSOFT International Symposium on Foundations of Software
  Engineering}} (Hong Kong, China) \emph{(\bibinfo{series}{FSE 2014})}.
  \bibinfo{publisher}{Association for Computing Machinery},
  \bibinfo{address}{New York, NY, USA}, \bibinfo{pages}{576–587}.
\newblock
\showISBNx{9781450330565}
\urldef\tempurl%
\url{https://doi.org/10.1145/2635868.2635869}
\showDOI{\tempurl}


\bibitem[Fu et~al\mbox{.}(2018)]%
        {10.1145/3168829}
\bibfield{author}{\bibinfo{person}{Xinwei Fu}, \bibinfo{person}{Dongyoon Lee},
  {and} \bibinfo{person}{Changhee Jung}.} \bibinfo{year}{2018}\natexlab{}.
\newblock \showarticletitle{NAdroid: Statically Detecting Ordering Violations
  in Android Applications}. In \bibinfo{booktitle}{\emph{Proceedings of the
  2018 International Symposium on Code Generation and Optimization}} (Vienna,
  Austria) \emph{(\bibinfo{series}{CGO 2018})}. \bibinfo{publisher}{Association
  for Computing Machinery}, \bibinfo{address}{New York, NY, USA},
  \bibinfo{pages}{62–74}.
\newblock
\showISBNx{9781450356176}
\urldef\tempurl%
\url{https://doi.org/10.1145/3168829}
\showDOI{\tempurl}


\bibitem[Google(2023a)]%
        {monkey}
\bibfield{author}{\bibinfo{person}{Google}.} \bibinfo{year}{2023}\natexlab{a}.
\newblock \bibinfo{title}{Android Monkey,
  \url{https://developer.android.com/studio/test/ monkey}}.
\newblock
\newblock
\newblock
\shownote{Accessed July 2023}.


\bibitem[Google(2023b)]%
        {flutter}
\bibfield{author}{\bibinfo{person}{Google}.} \bibinfo{year}{2023}\natexlab{b}.
\newblock \bibinfo{title}{Flutter}.
\newblock
\newblock
\urldef\tempurl%
\url{https://flutter.dev}
\showURL{%
\tempurl}
\newblock
\shownote{Accessed December 2023}.


\bibitem[Gordon et~al\mbox{.}(2015)]%
        {gordon2015information}
\bibfield{author}{\bibinfo{person}{Michael~I Gordon}, \bibinfo{person}{Deokhwan
  Kim}, \bibinfo{person}{Jeff~H Perkins}, \bibinfo{person}{Limei Gilham},
  \bibinfo{person}{Nguyen Nguyen}, {and} \bibinfo{person}{Martin~C Rinard}.}
  \bibinfo{year}{2015}\natexlab{}.
\newblock \showarticletitle{Information flow analysis of android applications
  in droidsafe.}. In \bibinfo{booktitle}{\emph{NDSS}},
  Vol.~\bibinfo{volume}{15}. \bibinfo{pages}{110}.
\newblock


\bibitem[Intelligence and Team(2022)]%
        {scylla}
\bibfield{author}{\bibinfo{person}{Satori~Threat Intelligence} {and}
  \bibinfo{person}{Research Team}.} \bibinfo{year}{2022}\natexlab{}.
\newblock \bibinfo{title}{Poseidon’s Offspring: Charybdis and Scylla,
  \url{https://www.humansecurity.com/learn/blog/poseidons-offspring-charybdis-and-scylla}}.
\newblock
\newblock
\newblock
\shownote{Accessed July 2023}.


\bibitem[Karakaya and Bodden(2021)]%
        {9610670}
\bibfield{author}{\bibinfo{person}{Kadiray Karakaya} {and}
  \bibinfo{person}{Eric Bodden}.} \bibinfo{year}{2021}\natexlab{}.
\newblock \showarticletitle{SootFX: A Static Code Feature Extraction Tool for
  Java and Android}. In \bibinfo{booktitle}{\emph{2021 IEEE 21st International
  Working Conference on Source Code Analysis and Manipulation (SCAM)}}.
  \bibinfo{pages}{181--186}.
\newblock
\urldef\tempurl%
\url{https://doi.org/10.1109/SCAM52516.2021.00030}
\showDOI{\tempurl}


\bibitem[Klieber et~al\mbox{.}(2014)]%
        {klieber2014android}
\bibfield{author}{\bibinfo{person}{William Klieber}, \bibinfo{person}{Lori
  Flynn}, \bibinfo{person}{Amar Bhosale}, \bibinfo{person}{Limin Jia}, {and}
  \bibinfo{person}{Lujo Bauer}.} \bibinfo{year}{2014}\natexlab{}.
\newblock \showarticletitle{Android taint flow analysis for app sets}. In
  \bibinfo{booktitle}{\emph{Proceedings of the 3rd ACM SIGPLAN International
  Workshop on the State of the Art in Java Program Analysis}}.
  \bibinfo{pages}{1--6}.
\newblock


\bibitem[Kuznetsov et~al\mbox{.}(2018)]%
        {10.1145/3197231.3197232}
\bibfield{author}{\bibinfo{person}{Konstantin Kuznetsov},
  \bibinfo{person}{Vitalii Avdiienko}, \bibinfo{person}{Alessandra Gorla},
  {and} \bibinfo{person}{Andreas Zeller}.} \bibinfo{year}{2018}\natexlab{}.
\newblock \showarticletitle{Analyzing the User Interface of Android Apps}. In
  \bibinfo{booktitle}{\emph{Proceedings of the 5th International Conference on
  Mobile Software Engineering and Systems}} (Gothenburg, Sweden)
  \emph{(\bibinfo{series}{MOBILESoft '18})}. \bibinfo{publisher}{Association
  for Computing Machinery}, \bibinfo{address}{New York, NY, USA},
  \bibinfo{pages}{84–87}.
\newblock
\showISBNx{9781450357128}
\urldef\tempurl%
\url{https://doi.org/10.1145/3197231.3197232}
\showDOI{\tempurl}


\bibitem[Lai and Rubin(2020)]%
        {10.1109/ASE.2019.00021}
\bibfield{author}{\bibinfo{person}{Duling Lai} {and} \bibinfo{person}{Julia
  Rubin}.} \bibinfo{year}{2020}\natexlab{}.
\newblock \showarticletitle{Goal-Driven Exploration for Android Applications}.
  In \bibinfo{booktitle}{\emph{Proceedings of the 34th IEEE/ACM International
  Conference on Automated Software Engineering}} (San Diego, California)
  \emph{(\bibinfo{series}{ASE '19})}. \bibinfo{publisher}{IEEE Press},
  \bibinfo{pages}{115–127}.
\newblock
\showISBNx{9781728125084}
\urldef\tempurl%
\url{https://doi.org/10.1109/ASE.2019.00021}
\showDOI{\tempurl}


\bibitem[Lee et~al\mbox{.}(2016)]%
        {10.1145/2970276.2970368}
\bibfield{author}{\bibinfo{person}{Sungho Lee}, \bibinfo{person}{Julian Dolby},
  {and} \bibinfo{person}{Sukyoung Ryu}.} \bibinfo{year}{2016}\natexlab{}.
\newblock \showarticletitle{HybriDroid: Static Analysis Framework for Android
  Hybrid Applications}. In \bibinfo{booktitle}{\emph{Proceedings of the 31st
  IEEE/ACM International Conference on Automated Software Engineering}}
  (Singapore, Singapore) \emph{(\bibinfo{series}{ASE '16})}.
  \bibinfo{publisher}{Association for Computing Machinery},
  \bibinfo{address}{New York, NY, USA}, \bibinfo{pages}{250–261}.
\newblock
\showISBNx{9781450338455}
\urldef\tempurl%
\url{https://doi.org/10.1145/2970276.2970368}
\showDOI{\tempurl}


\bibitem[Lee et~al\mbox{.}(2017)]%
        {10.1109/ICSE.2017.36}
\bibfield{author}{\bibinfo{person}{Youn~Kyu Lee}, \bibinfo{person}{Jae~young
  Bang}, \bibinfo{person}{Gholamreza Safi}, \bibinfo{person}{Arman Shahbazian},
  \bibinfo{person}{Yixue Zhao}, {and} \bibinfo{person}{Nenad Medvidovic}.}
  \bibinfo{year}{2017}\natexlab{}.
\newblock \showarticletitle{A SEALANT for Inter-App Security Holes in Android}.
  In \bibinfo{booktitle}{\emph{Proceedings of the 39th International Conference
  on Software Engineering}} (Buenos Aires, Argentina)
  \emph{(\bibinfo{series}{ICSE '17})}. \bibinfo{publisher}{IEEE Press},
  \bibinfo{pages}{312–323}.
\newblock
\showISBNx{9781538638682}
\urldef\tempurl%
\url{https://doi.org/10.1109/ICSE.2017.36}
\showDOI{\tempurl}


\bibitem[Li et~al\mbox{.}(2022)]%
        {10.1145/3540250.3549142}
\bibfield{author}{\bibinfo{person}{Chaoran Li}, \bibinfo{person}{Xiao Chen},
  \bibinfo{person}{Ruoxi Sun}, \bibinfo{person}{Minhui Xue},
  \bibinfo{person}{Sheng Wen}, \bibinfo{person}{Muhammad~Ejaz Ahmed},
  \bibinfo{person}{Seyit Camtepe}, {and} \bibinfo{person}{Yang Xiang}.}
  \bibinfo{year}{2022}\natexlab{}.
\newblock \showarticletitle{Cross-Language Android Permission Specification}.
  In \bibinfo{booktitle}{\emph{Proceedings of the 30th ACM Joint European
  Software Engineering Conference and Symposium on the Foundations of Software
  Engineering}} (Singapore, Singapore) \emph{(\bibinfo{series}{ESEC/FSE
  2022})}. \bibinfo{publisher}{Association for Computing Machinery},
  \bibinfo{address}{New York, NY, USA}, \bibinfo{pages}{772–783}.
\newblock
\showISBNx{9781450394130}
\urldef\tempurl%
\url{https://doi.org/10.1145/3540250.3549142}
\showDOI{\tempurl}


\bibitem[Li et~al\mbox{.}(2018)]%
        {8719450}
\bibfield{author}{\bibinfo{person}{Cong Li}, \bibinfo{person}{Chang Xu},
  \bibinfo{person}{Lili Wei}, \bibinfo{person}{Jue Wang}, \bibinfo{person}{Jun
  Ma}, {and} \bibinfo{person}{Jian Lu}.} \bibinfo{year}{2018}\natexlab{}.
\newblock \showarticletitle{ELEGANT: Towards Effective Location of
  Fragmentation-Induced Compatibility Issues for Android Apps}. In
  \bibinfo{booktitle}{\emph{2018 25th Asia-Pacific Software Engineering
  Conference (APSEC)}}. \bibinfo{pages}{278--287}.
\newblock
\urldef\tempurl%
\url{https://doi.org/10.1109/APSEC.2018.00042}
\showDOI{\tempurl}


\bibitem[Li et~al\mbox{.}(2015)]%
        {10.5555/2818754.2818791}
\bibfield{author}{\bibinfo{person}{Li Li}, \bibinfo{person}{Alexandre Bartel},
  \bibinfo{person}{Tegawend\'{e}~F. Bissyand\'{e}}, \bibinfo{person}{Jacques
  Klein}, \bibinfo{person}{Yves Le~Traon}, \bibinfo{person}{Steven Arzt},
  \bibinfo{person}{Siegfried Rasthofer}, \bibinfo{person}{Eric Bodden},
  \bibinfo{person}{Damien Octeau}, {and} \bibinfo{person}{Patrick McDaniel}.}
  \bibinfo{year}{2015}\natexlab{}.
\newblock \showarticletitle{IccTA: Detecting Inter-Component Privacy Leaks in
  Android Apps}. In \bibinfo{booktitle}{\emph{Proceedings of the 37th
  International Conference on Software Engineering - Volume 1}} (Florence,
  Italy) \emph{(\bibinfo{series}{ICSE '15})}. \bibinfo{publisher}{IEEE Press},
  \bibinfo{pages}{280–291}.
\newblock
\showISBNx{9781479919345}


\bibitem[Li et~al\mbox{.}(2016)]%
        {10.1145/2931037.2931044}
\bibfield{author}{\bibinfo{person}{Li Li}, \bibinfo{person}{Tegawend\'{e}~F.
  Bissyand\'{e}}, \bibinfo{person}{Damien Octeau}, {and}
  \bibinfo{person}{Jacques Klein}.} \bibinfo{year}{2016}\natexlab{}.
\newblock \showarticletitle{DroidRA: Taming Reflection to Support Whole-Program
  Analysis of Android Apps}. In \bibinfo{booktitle}{\emph{Proceedings of the
  25th International Symposium on Software Testing and Analysis}}
  (Saarbr\"{u}cken, Germany) \emph{(\bibinfo{series}{ISSTA 2016})}.
  \bibinfo{publisher}{Association for Computing Machinery},
  \bibinfo{address}{New York, NY, USA}, \bibinfo{pages}{318–329}.
\newblock
\showISBNx{9781450343909}
\urldef\tempurl%
\url{https://doi.org/10.1145/2931037.2931044}
\showDOI{\tempurl}


\bibitem[Mahmud et~al\mbox{.}(2022)]%
        {10.1145/3510454.3516854}
\bibfield{author}{\bibinfo{person}{Tarek Mahmud}, \bibinfo{person}{Meiru Che},
  {and} \bibinfo{person}{Guowei Yang}.} \bibinfo{year}{2022}\natexlab{}.
\newblock \showarticletitle{ACID: An API Compatibility Issue Detector for
  Android Apps}. In \bibinfo{booktitle}{\emph{Proceedings of the ACM/IEEE 44th
  International Conference on Software Engineering: Companion Proceedings}}
  (Pittsburgh, Pennsylvania) \emph{(\bibinfo{series}{ICSE '22})}.
  \bibinfo{publisher}{Association for Computing Machinery},
  \bibinfo{address}{New York, NY, USA}, \bibinfo{pages}{1–5}.
\newblock
\showISBNx{9781450392235}
\urldef\tempurl%
\url{https://doi.org/10.1145/3510454.3516854}
\showDOI{\tempurl}


\bibitem[Mann and Whitney(1947)]%
        {10.1214/aoms/1177730491}
\bibfield{author}{\bibinfo{person}{H.~B. Mann} {and} \bibinfo{person}{D.~R.
  Whitney}.} \bibinfo{year}{1947}\natexlab{}.
\newblock \showarticletitle{On a Test of Whether one of Two Random Variables is
  Stochastically Larger than the Other}.
\newblock \bibinfo{journal}{\emph{Annals of Mathematical Statistics}}
  \bibinfo{volume}{18}, \bibinfo{number}{1} (\bibinfo{date}{03}
  \bibinfo{year}{1947}), \bibinfo{pages}{50--60}.
\newblock
\urldef\tempurl%
\url{https://doi.org/10.1214/aoms/1177730491}
\showDOI{\tempurl}


\bibitem[Octeau et~al\mbox{.}(2015)]%
        {10.5555/2818754.2818767}
\bibfield{author}{\bibinfo{person}{Damien Octeau}, \bibinfo{person}{Daniel
  Luchaup}, \bibinfo{person}{Matthew Dering}, \bibinfo{person}{Somesh Jha},
  {and} \bibinfo{person}{Patrick McDaniel}.} \bibinfo{year}{2015}\natexlab{}.
\newblock \showarticletitle{Composite Constant Propagation: Application to
  Android Inter-Component Communication Analysis}. In
  \bibinfo{booktitle}{\emph{Proceedings of the 37th International Conference on
  Software Engineering - Volume 1}} (Florence, Italy)
  \emph{(\bibinfo{series}{ICSE '15})}. \bibinfo{publisher}{IEEE Press},
  \bibinfo{pages}{77–88}.
\newblock
\showISBNx{9781479919345}


\bibitem[Onwuzurike et~al\mbox{.}(2019)]%
        {10.1145/3313391}
\bibfield{author}{\bibinfo{person}{Lucky Onwuzurike}, \bibinfo{person}{Enrico
  Mariconti}, \bibinfo{person}{Panagiotis Andriotis},
  \bibinfo{person}{Emiliano~De Cristofaro}, \bibinfo{person}{Gordon Ross},
  {and} \bibinfo{person}{Gianluca Stringhini}.}
  \bibinfo{year}{2019}\natexlab{}.
\newblock \showarticletitle{MaMaDroid: Detecting Android Malware by Building
  Markov Chains of Behavioral Models (Extended Version)}.
\newblock \bibinfo{journal}{\emph{ACM Trans. Priv. Secur.}}
  \bibinfo{volume}{22}, \bibinfo{number}{2}, Article \bibinfo{articleno}{14}
  (\bibinfo{date}{apr} \bibinfo{year}{2019}), \bibinfo{numpages}{34}~pages.
\newblock
\showISSN{2471-2566}
\urldef\tempurl%
\url{https://doi.org/10.1145/3313391}
\showDOI{\tempurl}


\bibitem[Pauck and Wehrheim(2021)]%
        {9610738}
\bibfield{author}{\bibinfo{person}{Felix Pauck} {and} \bibinfo{person}{Heike
  Wehrheim}.} \bibinfo{year}{2021}\natexlab{}.
\newblock \showarticletitle{Jicer: Simplifying Cooperative Android App Analysis
  Tasks}. In \bibinfo{booktitle}{\emph{2021 IEEE 21st International Working
  Conference on Source Code Analysis and Manipulation (SCAM)}}.
  \bibinfo{pages}{187--197}.
\newblock
\urldef\tempurl%
\url{https://doi.org/10.1109/SCAM52516.2021.00031}
\showDOI{\tempurl}


\bibitem[Reif et~al\mbox{.}(2019)]%
        {10.1145/3293882.3330555}
\bibfield{author}{\bibinfo{person}{Michael Reif}, \bibinfo{person}{Florian
  K\"{u}bler}, \bibinfo{person}{Michael Eichberg}, \bibinfo{person}{Dominik
  Helm}, {and} \bibinfo{person}{Mira Mezini}.} \bibinfo{year}{2019}\natexlab{}.
\newblock \showarticletitle{Judge: Identifying, Understanding, and Evaluating
  Sources of Unsoundness in Call Graphs}. In
  \bibinfo{booktitle}{\emph{Proceedings of the 28th ACM SIGSOFT International
  Symposium on Software Testing and Analysis}} (Beijing, China)
  \emph{(\bibinfo{series}{ISSTA 2019})}. \bibinfo{publisher}{Association for
  Computing Machinery}, \bibinfo{address}{New York, NY, USA},
  \bibinfo{pages}{251–261}.
\newblock
\showISBNx{9781450362245}
\urldef\tempurl%
\url{https://doi.org/10.1145/3293882.3330555}
\showDOI{\tempurl}


\bibitem[repository(2023)]%
        {columbusrepository}
\bibfield{author}{\bibinfo{person}{Columbus repository}.}
  \bibinfo{year}{2023}\natexlab{}.
\newblock \bibinfo{title}{\url{https://github.com/ucsb-seclab/columbus}}.
\newblock
\newblock
\newblock
\shownote{Accessed December 2023}.


\bibitem[Samhi(2023)]%
        {jordan_samhi_thesis_2023}
\bibfield{author}{\bibinfo{person}{Jordan Samhi}.}
  \bibinfo{year}{2023}\natexlab{}.
\newblock \emph{\bibinfo{title}{Analyzing the Unanalyzable: an Application to
  Android Apps}}.
\newblock \bibinfo{thesistype}{Ph.\,D. Dissertation}.
  \bibinfo{school}{University of Luxembourg, Luxembourg City, Luxembourg}.
\newblock
\urldef\tempurl%
\url{http://hdl.handle.net/10993/54372}
\showURL{%
\tempurl}


\bibitem[Samhi et~al\mbox{.}(2021)]%
        {10.1109/ICSE43902.2021.00126}
\bibfield{author}{\bibinfo{person}{J. Samhi}, \bibinfo{person}{A. Bartel},
  \bibinfo{person}{T.~F. Bissyande}, {and} \bibinfo{person}{J. Klein}.}
  \bibinfo{year}{2021}\natexlab{}.
\newblock \showarticletitle{RAICC: Revealing Atypical Inter-Component
  Communication in Android Apps}. In \bibinfo{booktitle}{\emph{2021 IEEE/ACM
  43rd International Conference on Software Engineering (ICSE)}}.
  \bibinfo{publisher}{IEEE Computer Society}, \bibinfo{address}{Los Alamitos,
  CA, USA}, \bibinfo{pages}{1398--1409}.
\newblock
\showISBNx{9781450390859}
\urldef\tempurl%
\url{https://doi.org/10.1109/ICSE43902.2021.00126}
\showDOI{\tempurl}


\bibitem[Samhi et~al\mbox{.}(2024)]%
        {androlibzoo}
\bibfield{author}{\bibinfo{person}{J. Samhi}, \bibinfo{person}{T.~F.
  Bissyande}, {and} \bibinfo{person}{J. Klein}.}
  \bibinfo{year}{2024}\natexlab{}.
\newblock \showarticletitle{AndroLibZoo: A Reliable Dataset of Libraries Based
  on Software Dependency Analysis}. In \bibinfo{booktitle}{\emph{2024 IEEE/ACM
  21st International Conference on Mining Software Repositories (MSR)}}.
\newblock


\bibitem[Samhi et~al\mbox{.}(2022a)]%
        {10.1145/3510003.3512766}
\bibfield{author}{\bibinfo{person}{Jordan Samhi}, \bibinfo{person}{Jun Gao},
  \bibinfo{person}{Nadia Daoudi}, \bibinfo{person}{Pierre Graux},
  \bibinfo{person}{Henri Hoyez}, \bibinfo{person}{Xiaoyu Sun},
  \bibinfo{person}{Kevin Allix}, \bibinfo{person}{Tegawend{\'e}~F
  Bissyand{\'e}}, {and} \bibinfo{person}{Jacques Klein}.}
  \bibinfo{year}{2022}\natexlab{a}.
\newblock \showarticletitle{JuCify: A Step Towards Android Code Unification for
  Enhanced Static Analysis}. In \bibinfo{booktitle}{\emph{2022 IEEE/ACM 44th
  International Conference on Software Engineering (ICSE)}}.
  \bibinfo{publisher}{IEEE Computer Society}, \bibinfo{address}{Los Alamitos,
  CA, USA}, \bibinfo{pages}{1232--1244}.
\newblock
\urldef\tempurl%
\url{https://doi.org/10.1145/3510003.3512766}
\showDOI{\tempurl}


\bibitem[Samhi et~al\mbox{.}(2022b)]%
        {10.1145/3510003.3510135}
\bibfield{author}{\bibinfo{person}{J. Samhi}, \bibinfo{person}{L. Li},
  \bibinfo{person}{T.~F. Bissyande}, {and} \bibinfo{person}{J. Klein}.}
  \bibinfo{year}{2022}\natexlab{b}.
\newblock \showarticletitle{Difuzer: Uncovering Suspicious Hidden Sensitive
  Operations in Android Apps}. In \bibinfo{booktitle}{\emph{2022 IEEE/ACM 44th
  International Conference on Software Engineering (ICSE)}}.
  \bibinfo{publisher}{IEEE Computer Society}, \bibinfo{address}{Los Alamitos,
  CA, USA}, \bibinfo{pages}{723--735}.
\newblock
\urldef\tempurl%
\url{https://doi.org/10.1145/3510003.3510135}
\showDOI{\tempurl}


\bibitem[Service(2023)]%
        {audioservice}
\bibfield{author}{\bibinfo{person}{Audio Service}.}
  \bibinfo{year}{2023}\natexlab{}.
\newblock \bibinfo{title}{\url{https://github.com/ryanheise/audio_service}}.
\newblock
\newblock
\newblock
\shownote{Accessed December 2023}.


\bibitem[Sui et~al\mbox{.}(2018)]%
        {10.1007/978-3-030-02768-1_4}
\bibfield{author}{\bibinfo{person}{Li Sui}, \bibinfo{person}{Jens Dietrich},
  \bibinfo{person}{Michael Emery}, \bibinfo{person}{Shawn Rasheed}, {and}
  \bibinfo{person}{Amjed Tahir}.} \bibinfo{year}{2018}\natexlab{}.
\newblock \showarticletitle{On the Soundness of Call Graph Construction in the
  Presence of Dynamic Language Features - A Benchmark and Tool Evaluation}. In
  \bibinfo{booktitle}{\emph{Programming Languages and Systems}},
  \bibfield{editor}{\bibinfo{person}{Sukyoung Ryu}} (Ed.).
  \bibinfo{publisher}{Springer International Publishing},
  \bibinfo{address}{Cham}, \bibinfo{pages}{69--88}.
\newblock
\showISBNx{978-3-030-02768-1}


\bibitem[Unity(2023)]%
        {unity}
\bibfield{author}{\bibinfo{person}{Unity}.} \bibinfo{year}{2023}\natexlab{}.
\newblock \bibinfo{title}{Unity}.
\newblock
\newblock
\urldef\tempurl%
\url{https://unity.com}
\showURL{%
\tempurl}
\newblock
\shownote{Accessed December 2023}.


\bibitem[Wang et~al\mbox{.}(2021)]%
        {10.1145/3460319.3464828}
\bibfield{author}{\bibinfo{person}{Wenyu Wang}, \bibinfo{person}{Wing Lam},
  {and} \bibinfo{person}{Tao Xie}.} \bibinfo{year}{2021}\natexlab{}.
\newblock \showarticletitle{An Infrastructure Approach to Improving
  Effectiveness of Android UI Testing Tools}. In
  \bibinfo{booktitle}{\emph{Proceedings of the 30th ACM SIGSOFT International
  Symposium on Software Testing and Analysis}} (Virtual, Denmark)
  \emph{(\bibinfo{series}{ISSTA 2021})}. \bibinfo{publisher}{Association for
  Computing Machinery}, \bibinfo{address}{New York, NY, USA},
  \bibinfo{pages}{165–176}.
\newblock
\showISBNx{9781450384599}
\urldef\tempurl%
\url{https://doi.org/10.1145/3460319.3464828}
\showDOI{\tempurl}


\bibitem[Wang et~al\mbox{.}(2018)]%
        {9000056}
\bibfield{author}{\bibinfo{person}{Wenyu Wang}, \bibinfo{person}{Dengfeng Li},
  \bibinfo{person}{Wei Yang}, \bibinfo{person}{Yurui Cao},
  \bibinfo{person}{Zhenwen Zhang}, \bibinfo{person}{Yuetang Deng}, {and}
  \bibinfo{person}{Tao Xie}.} \bibinfo{year}{2018}\natexlab{}.
\newblock \showarticletitle{An Empirical Study of Android Test Generation Tools
  in Industrial Cases}. In \bibinfo{booktitle}{\emph{2018 33rd IEEE/ACM
  International Conference on Automated Software Engineering (ASE)}}.
  \bibinfo{pages}{738--748}.
\newblock
\urldef\tempurl%
\url{https://doi.org/10.1145/3238147.3240465}
\showDOI{\tempurl}


\bibitem[Wang et~al\mbox{.}(2016)]%
        {10.1145/2931021.2931026}
\bibfield{author}{\bibinfo{person}{Yan Wang}, \bibinfo{person}{Hailong Zhang},
  {and} \bibinfo{person}{Atanas Rountev}.} \bibinfo{year}{2016}\natexlab{}.
\newblock \showarticletitle{On the Unsoundness of Static Analysis for Android
  GUIs}. In \bibinfo{booktitle}{\emph{Proceedings of the 5th ACM SIGPLAN
  International Workshop on State Of the Art in Program Analysis}} (Santa
  Barbara, CA, USA) \emph{(\bibinfo{series}{SOAP 2016})}.
  \bibinfo{publisher}{Association for Computing Machinery},
  \bibinfo{address}{New York, NY, USA}, \bibinfo{pages}{18–23}.
\newblock
\showISBNx{9781450343855}
\urldef\tempurl%
\url{https://doi.org/10.1145/2931021.2931026}
\showDOI{\tempurl}


\bibitem[Wei et~al\mbox{.}(2014)]%
        {10.1145/2660267.2660357}
\bibfield{author}{\bibinfo{person}{Fengguo Wei}, \bibinfo{person}{Sankardas
  Roy}, \bibinfo{person}{Xinming Ou}, {and} \bibinfo{person}{Robby}.}
  \bibinfo{year}{2014}\natexlab{}.
\newblock \showarticletitle{Amandroid: A Precise and General Inter-Component
  Data Flow Analysis Framework for Security Vetting of Android Apps}. In
  \bibinfo{booktitle}{\emph{Proceedings of the 2014 ACM SIGSAC Conference on
  Computer and Communications Security}} (Scottsdale, Arizona, USA)
  \emph{(\bibinfo{series}{CCS '14})}. \bibinfo{publisher}{Association for
  Computing Machinery}, \bibinfo{address}{New York, NY, USA},
  \bibinfo{pages}{1329–1341}.
\newblock
\showISBNx{9781450329576}
\urldef\tempurl%
\url{https://doi.org/10.1145/2660267.2660357}
\showDOI{\tempurl}


\bibitem[Wu et~al\mbox{.}(2021)]%
        {9505167}
\bibfield{author}{\bibinfo{person}{Daoyuan Wu}, \bibinfo{person}{Debin Gao},
  \bibinfo{person}{Robert~H. Deng}, {and} \bibinfo{person}{Chang Rocky K.~C.}}
  \bibinfo{year}{2021}\natexlab{}.
\newblock \showarticletitle{When Program Analysis Meets Bytecode Search:
  Targeted and Efficient Inter-procedural Analysis of Modern Android Apps in
  BackDroid}. In \bibinfo{booktitle}{\emph{2021 51st Annual IEEE/IFIP
  International Conference on Dependable Systems and Networks (DSN)}}.
  \bibinfo{pages}{543--554}.
\newblock
\urldef\tempurl%
\url{https://doi.org/10.1109/DSN48987.2021.00063}
\showDOI{\tempurl}


\bibitem[Wu et~al\mbox{.}(2016)]%
        {10.1109/TSE.2016.2547385}
\bibfield{author}{\bibinfo{person}{Tianyong Wu}, \bibinfo{person}{Jierui Liu},
  \bibinfo{person}{Zhenbo Xu}, \bibinfo{person}{Chaorong Guo},
  \bibinfo{person}{Yanli Zhang}, \bibinfo{person}{Jun Yan}, {and}
  \bibinfo{person}{Jian Zhang}.} \bibinfo{year}{2016}\natexlab{}.
\newblock \showarticletitle{Light-Weight, Inter-Procedural and Callback-Aware
  Resource Leak Detection for Android Apps}.
\newblock \bibinfo{journal}{\emph{IEEE Transactions on Software Engineering}}
  \bibinfo{volume}{42}, \bibinfo{number}{11} (\bibinfo{year}{2016}),
  \bibinfo{pages}{1054--1076}.
\newblock
\urldef\tempurl%
\url{https://doi.org/10.1109/TSE.2016.2547385}
\showDOI{\tempurl}


\bibitem[XAMARIN(2023)]%
        {xamarin}
\bibfield{author}{\bibinfo{person}{XAMARIN}.} \bibinfo{year}{2023}\natexlab{}.
\newblock \bibinfo{title}{\url{https://dotnet.microsoft.com/apps/xamarin}}.
\newblock
\newblock
\newblock
\shownote{Accessed December 2023}.


\bibitem[Yan et~al\mbox{.}(2022)]%
        {10.1109/ICSE-Companion55297.2022.9793791}
\bibfield{author}{\bibinfo{person}{Jiwei Yan}, \bibinfo{person}{Shixin Zhang},
  \bibinfo{person}{Yepang Liu}, \bibinfo{person}{Jun Yan}, {and}
  \bibinfo{person}{Jian Zhang}.} \bibinfo{year}{2022}\natexlab{}.
\newblock \showarticletitle{ICCBot: Fragment-Aware and Context-Sensitive ICC
  Resolution for Android Applications}. In \bibinfo{booktitle}{\emph{2022
  IEEE/ACM 44th International Conference on Software Engineering: Companion
  Proceedings (ICSE-Companion)}}. \bibinfo{pages}{105--109}.
\newblock
\urldef\tempurl%
\url{https://doi.org/10.1109/ICSE-Companion55297.2022.9793791}
\showDOI{\tempurl}


\bibitem[Yang et~al\mbox{.}(2015a)]%
        {7194564}
\bibfield{author}{\bibinfo{person}{Shengqian Yang}, \bibinfo{person}{Dacong
  Yan}, \bibinfo{person}{Haowei Wu}, \bibinfo{person}{Yan Wang}, {and}
  \bibinfo{person}{Atanas Rountev}.} \bibinfo{year}{2015}\natexlab{a}.
\newblock \showarticletitle{Static Control-Flow Analysis of User-Driven
  Callbacks in Android Applications}. In \bibinfo{booktitle}{\emph{2015
  IEEE/ACM 37th IEEE International Conference on Software Engineering}},
  Vol.~\bibinfo{volume}{1}. \bibinfo{pages}{89--99}.
\newblock
\urldef\tempurl%
\url{https://doi.org/10.1109/ICSE.2015.31}
\showDOI{\tempurl}


\bibitem[Yang et~al\mbox{.}(2015b)]%
        {10.1109/ICSE.2015.31}
\bibfield{author}{\bibinfo{person}{Shengqian Yang}, \bibinfo{person}{Dacong
  Yan}, \bibinfo{person}{Haowei Wu}, \bibinfo{person}{Yan Wang}, {and}
  \bibinfo{person}{Atanas Rountev}.} \bibinfo{year}{2015}\natexlab{b}.
\newblock \showarticletitle{Static Control-Flow Analysis of User-Driven
  Callbacks in Android Applications}. In \bibinfo{booktitle}{\emph{2015
  IEEE/ACM 37th IEEE International Conference on Software Engineering}},
  Vol.~\bibinfo{volume}{1}. \bibinfo{pages}{89--99}.
\newblock
\urldef\tempurl%
\url{https://doi.org/10.1109/ICSE.2015.31}
\showDOI{\tempurl}


\end{thebibliography}

\end{document}